\documentclass[lettersize,journal]{IEEEtran}
\usepackage{amsmath, amsfonts}
\usepackage{array}
\usepackage{textcomp}
\usepackage{stfloats}
\usepackage{url}
\usepackage{verbatim}
\usepackage{graphicx}
\usepackage{tabularx}
\usepackage{booktabs}
\usepackage{algorithm}
\usepackage{multirow}
\usepackage{orcidlink}
\usepackage{algpseudocode}

\def\BibTeX{{\rm B\kern-.05em{\sc i\kern-.025em b}\kern-.08em
    T\kern-.1667em\lower.7ex\hbox{E}\kern-.125emX}}
\usepackage{balance}

\begin{document}
\title{A Differentiable Material Point Method Framework\\for Shape Morphing}

\author{M.~Xu\raisebox{0.5ex}{\orcidlink{0009-0005-2977-7993}}, 
        C.~Y.~Song\raisebox{0.5ex}{\orcidlink{0000-0002-2495-1383}}, 
        D.~Levin\raisebox{0.5ex}{\orcidlink{0000-0001-7079-1934}}, 
        and~D.~Hyde\raisebox{0.5ex}{\orcidlink{0009-0004-4950-5533}},~\IEEEmembership{Member,~IEEE}%
        
\thanks{M. Xu is with Simon Fraser University, Canada.}%
\thanks{D. Levin is with the University of Toronto, Canada.}%
\thanks{C. Y. Song and D. Hyde are with Vanderbilt University, USA.}%
\thanks{M. Xu and C. Y. Song contributed equally to this work.}%
}

\markboth{Journal of \LaTeX\ Class Files,~Vol.~18, No.~9, September~2020}%
{How to Use the IEEEtran \LaTeX \ Templates}

\maketitle

\begin{abstract}
We present a novel, physically-based morphing technique for elastic shapes, leveraging the differentiable material point method (MPM) with space-time control through per-particle deformation gradients to accommodate complex topology changes. This approach, grounded in MPM's natural handling of dynamic topologies, is enhanced by a chained iterative optimization technique, allowing for the creation of both succinct and extended morphing sequences that maintain coherence over time. Demonstrated across various challenging scenarios, our method is able to produce detailed elastic deformation and topology transitions, all grounded within our physics-based simulation framework.
\end{abstract}

\begin{IEEEkeywords}
Animation, Physical simulation.
\end{IEEEkeywords}

\section{Introduction}
\IEEEPARstart{K}{eyframe} control and shape interpolation have long been cornerstone techniques in the animation industry, offering animators the ability to craft sequences where characters and objects transition smoothly from one form to another \cite{gomes1999warping}.
    In {3D} computer animation, this effect, familiar to audiences worldwide, has seen blockbuster success as far back as \textit{Terminator 2} (1991) and \textit{The Abyss} (1989).
    These early applications captivated audiences with seamless character transitions.
    In more recent times, films like Disney's \textit{Moana} (2016) have pushed the boundaries further by integrating physically-based shape control to endow bodies like the ocean with a sense of personality and life, showcasing the evolution of morphing from a mere visual trick to a storytelling tool.
    
    Traditionally, shape morphing algorithms have been grounded in optimization.
    These algorithms are tasked with minimizing an objective that accounts for both the constraints imposed by keyframed shapes and additional terms designed to guide the morphing behavior.
    These supplementary terms are pivotal as they aim to imbue the morph with a layer of physical realism, whether through mechanisms like optimal transport \cite{doi:10.1137/22M1524254, 10.1145/2766963} or principles akin to Newton's second law \cite{terzopoulos1988deformable}.
    In this field, we have observed a dichotomy in the approach to these algorithms: on the one hand, methods drawn from fluidic physical models \cite{treuille2003keyframe, 2004-FluidControlAdjointMethod, stomakhin2017fluxed} are particularly adept at handling changes in topology using external forces such as wind. On the other hand, elasticity-based methods \cite{2012-DeformableObjectsAlive, min2019softcon} provide visually striking deformations by employing internal forces, but they typically assume a constant topology \cite{raveendran2012controlling}.
    In this paper, we integrate the properties of both elasticity-based and fluidic methods: our method relies on internal forces, typical of elasticity-based techniques, while naturally supporting topological changes seen in fluidic models to enable highly dynamic morphs. 
    
    We introduce a physically-based shape morphing algorithm tailored for elastic materials, with the goal of simultaneously achieving large deformations and dynamic topology changes throughout a morph.
    At the heart of our approach is the application of a differentiable material-point method ({MPM}) simulator, optimized for elastic materials and steered through deformation gradient control. Complementing this is a multi-pass optimization framework, designed for the efficient resolution of the complex shape-morphing and space-time challenges that arise.
    Additionally, our formulation includes a differentiable Eulerian loss function, a novel contribution that facilitates unique topological transformations.
    We showcase the advantages of our method with results that are characterized not only by the expected elastic-like deformations but also by dramatic topology changes.

\section{Related Work}
    Most shape morphing algorithms use optimization techniques to reproduce the shape trajectories defined by artists between keyframes.
    The distinctive characteristics and efficiency of each algorithm are determined by specifics of the given optimization problem, including the energy function that needs to be minimized, constraints applied to maintain the realism or style of the animation, and the mathematical discretization methods employed to solve these optimization problems.

    \textit{Shape morphing.} Shape morphing algorithms have been built upon various geometric representations.
    For instance, \cite{turk2005shape,breen2001level} demonstrated algorithms for implicit surfaces.
    \cite{bansal2018lie} presented a shape morphing/interpolation technique based on Lie body representations of triangular meshes.
    \cite{kilian2007geometric} recast shapes as points in a shape space endowed with Riemannian metrics that naturally enable tasks like shape morphing.
    Similar to our desire for physically-based morphing, \cite{bao2005physically} used point samples of a surface mesh to produce a method that yields more physically plausible results than linear blending.
    \cite{muller2004point} leveraged point sampling for both surfaces and volumes, and subsequently presented methods for shape morphing of elastoplastic objects; this work is similar in several ways to the present manuscript, although we use a physical simulation technique ({MPM}) that leads to points in our method, and we leverage differentiable simulation and various other optimization and control features not present in \cite{muller2004point}.

    Several other recent techniques are worth noting.
    For instance, \cite{dobashi2015simple} enabled morphing smoke, which can often exhibit complex geometric behavior.
    In another fluid application, \cite{raveendran2012controlling} demonstrated visually impressive techniques for controlling liquids, including morphing them per artist directions.
    Finally, we mention \cite{ludwig20153d}, which presented an unstructured lumigraph morphing algorithm that enables the simultaneous morphing of objects and textures.
    These examples demonstrate the diversity of methods for and applications of shape morphing.

    We also highlight the popular As-Rigid-As-Possible (ARAP) energy \cite{2000-ARAPShapeInterpolation} and its derivatives.
    ARAP is useful for shape interpolation tasks and involves the surface of an object represented by triangular or tetrahedral meshes that are deformed through the minimization of the ARAP energy.
    This approach can be extended using theories of deformation and deformation energy that emphasize physical realism \cite{2004-ActualMorphing, 2020-HamiltonianShapeInterpolation}, enhancing the accuracy and visual appeal of shape interpolation through more realistic energy models.
    Xu et al.\ \cite{2005-PoissonShapeInterpolation} proposed a new method for shape interpolation by solving the Poisson equation on the domain mesh, considering both vertex coordinates and surface orientations to prevent shrinkage issues and to generate physically plausible and visually satisfying deformation sequences. This represented a significant advancement in shape interpolation technology, greatly expanding its application possibilities in animation and visual effects.
    Liu et al. \cite{2011-ARAPSurfaceMorphing} reduced reliance on consistent tetrahedral meshes found in traditional shape interpolation approaches, instead utilizing triangular surface meshes to broaden the scope of shape interpolation. Their ``As-Rigid-As-Possible'' approach, integrating transformation vectors into the energy function, allowed for more efficient and flexible control of mesh deformation. This method makes ARAP-style approaches more accessible to modelers. 

    Recently, researchers have begun exploring the blending of shape morphing, deep learning, and data.
    Gao et al. \cite{2013-DataDrivenShapeMorphing} introduced a data-driven approach to shape interpolation, forming a local shape space for plausible deformation using a model database and recasting the deformation problem as a global optimization problem.
    Shape deformation and other geometry tasks were also achieved using geometry represented with neural fields \cite{yang2021geometry}.
    Given the increasingly rich interplays between learning and simulation (see e.g.\ \cite{sharpdeep}), we expect that our method---which is based on physical simulation---could also be enhanced with learning-based components in the future.

    \textit{Optimization and control.} Shape morphing and spatiotemporal optimization \cite{1988-SpacetimeConstraints} belong to very similar problem domains, with our research being closely related to the latter. In that vein of work,
    McNamara et al. \cite{2004-FluidControlAdjointMethod} proposed controlling fluid simulations using a keyframe-based loss function, minimized by calculating gradients through the adjoint method to align the fluid's motion with desired keyframes. Coros et al. \cite{2012-DeformableObjectsAlive} minimized a nonlinear time-dependent loss function using Sequential Quadratic Programming (SQP) to control mesh-based characters, addressing issues related to rest-state adaptation. Zhang et al. \cite{2020-ManipulateAmorphousMaterialsRL} explored a new method for controlling amorphous materials through externally controlled rigid contact bodies, presenting a way to learn a general controller. This research utilized reinforcement learning to innovate in manipulating physical materials.

   \begin{figure*}[!ht]
        \centering
        \includegraphics[width=\textwidth]{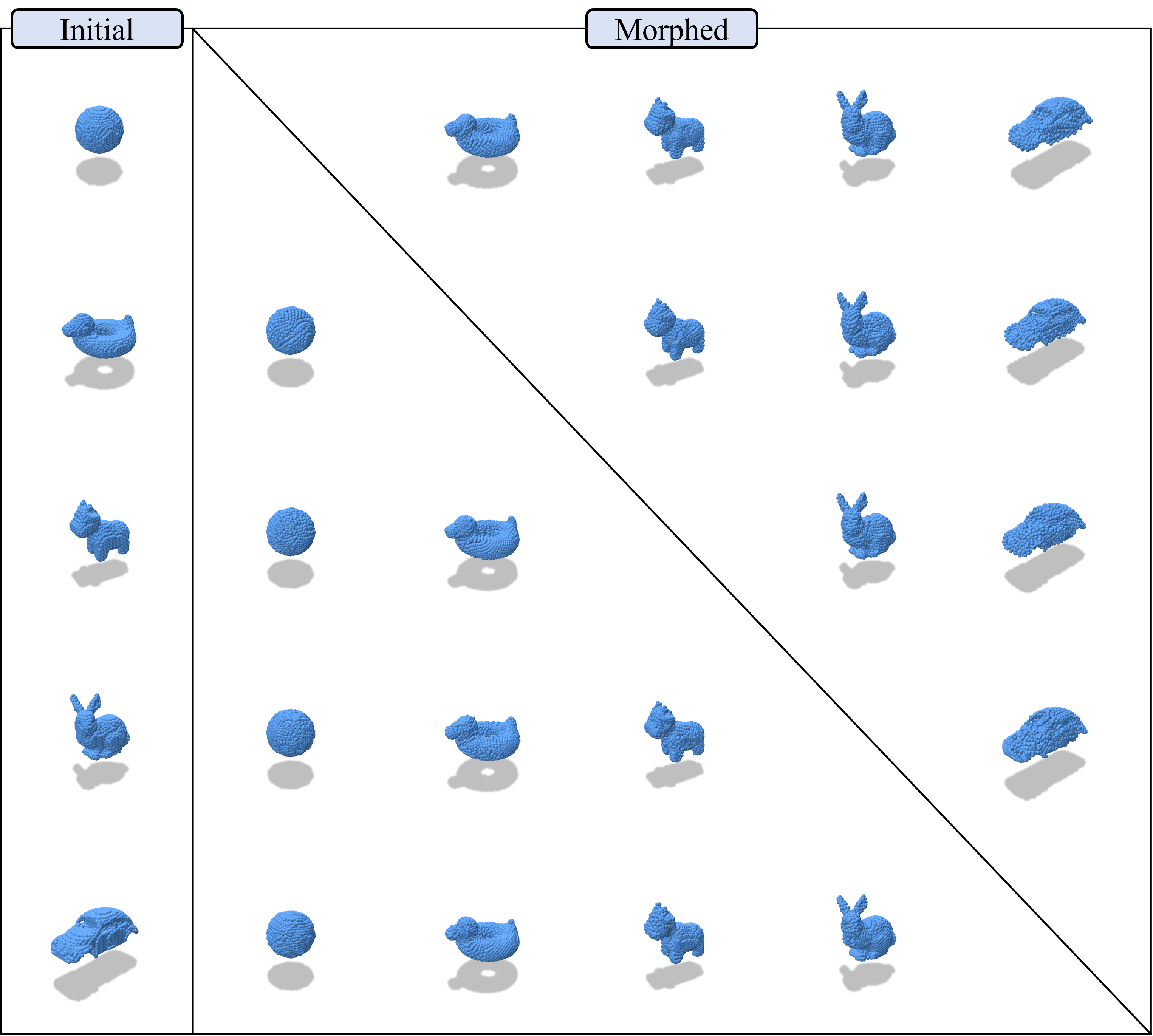}
        \caption{Morphing between pairs of five example meshes. Our test meshes exhibit different topologies (e.g., the inflatable duck) as well as fine-scale features (e.g., the ears on the bunny and cow). Our method successfully morphs initial geometries to their targets in all cases; we refer to the supplementary video for animations of these morphs.} 
        \label{fig:Overall_deform}
    \end{figure*}

    \textit{Differentiable simulation.} Our physically-based approach is built upon a differentiable simulation framework. Differentiable simulation (see e.g.\ \cite{coros2021differentiable}), due to its flexibility and efficiency, has attracted attention within the learning and graphics communities. Its core lies in the ability to calculate gradients through backpropagation, inspired by neural networks \cite{2020-DiffPhysicsSim}. It has been applied to tasks like soft robot control \cite{2018-ChainQueen, 2019-DiffTaichi}, material parameter estimation \cite{2019-DiffCloth, 2022-DiffSoftMultiBody}, and accelerating reinforcement learning \cite{2021-PODS, 2022-DiffSimRL}. Among other discretizations, the material point method \cite{sulsky1994particle,jiang2016material} has enjoyed differentiable formulations and implementations, e.g. \cite{2018-ChainQueen}, which we leverage in the present work.
    
    Our research builds on these various studies, including utilizing algorithms like {MLS-MPM} \cite{2018-MLSMPM}, to extend the applicability of differentiable simulation to additional complex geometry tasks.
    Similar to Hu et al. \cite{2018-ChainQueen}, who applied actuated stress control in soft body locomotion using differentiable {MLS-MPM}, our methodology proposes an algorithm suited for shape morphing requiring complex and topologically aware mesh configurations without needing intricate mesh setups as suggested by \cite{10.1145/3130800.3130820}.
    Our approach to optimizing and controlling complex dynamic phenomena like shape morphing is a novel combination of simulation and animation techniques that, as seen in the results, enables smooth deformations and dynamic topological changes.

\section{Simulation Overview}
    The basis for the present work is the material point method ({MPM}),
    which is a particle-in-cell method for simulating continuum materials.
    {MPM}'s hybrid particle-grid formulation requires no fixed connectivity between the discretized elements of the material domain (the material points).
    This naturally allows for large topology-changing deformations when compared to mesh-based methods like the finite element method. 
    Since its discovery 30 years ago \cite{sulsky1994particle}, various formulations and improvements have been made to the original {MPM}, e.g. \cite{2018-MLSMPM,2019-cdmpm,fang2020iq,bardenhagen2004generalized,solowski2021material,de2020material}.
    To provide background, we review the {MLS-MPM} variant that we use in this work \cite{2018-MLSMPM}. 
    
        \begin{figure*}
            \centering
            \includegraphics[width=\textwidth]{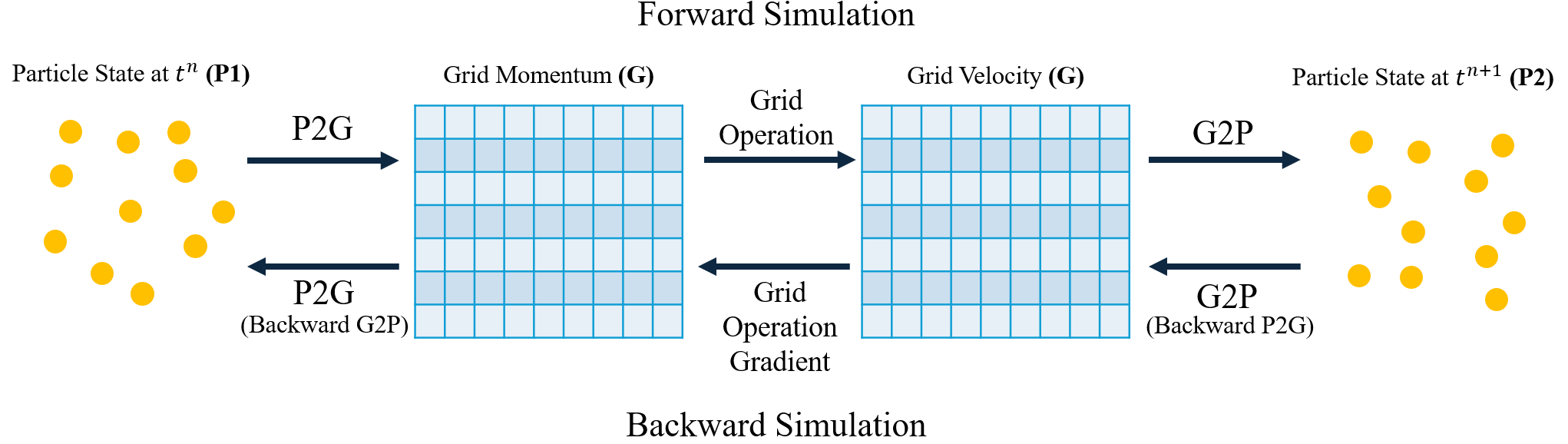}
            \caption{Simulation overview. \textmd{A timestep in {MPM} consists of five major computations: $\textbf{P1} \rightarrow \textbf{P2G} \rightarrow \textbf{G} \rightarrow \textbf{G2P} \rightarrow \textbf{P2}$. Particles store their own physical properties but interact with each other through the background simulation grid. }}
            \label{fig:MLS_MPM}
        \end{figure*}
        
    The method starts with a continuum discretized into material points that carry physical properties of its domain, such as mass $m_p$, volume $V_p$, position $\mathbf{x}_p$, velocity $\mathbf{v}_p$, and deformation gradient $\mathbf{F}_p$.
    We use APIC transfer kernels \cite{jiang2017angular}, which further requires an affine matrix $\mathbf{C}_p$ to be stored on the particles.
    
    In MPM, updates to particle states are generally computed on a background grid.
    On each timestep, the background grid interpolates material point quantities like velocity to grid nodes, computes updates for quantities like momentum, then interpolates updated quantities back to the particles.
    The use of a regular background grid enables using efficient, scalable solvers that have been optimized for grids.
    
    Each timestep of the {MLS-MPM} algorithm we implement can be split into five major steps, as seen in Figure \ref{fig:MLS_MPM}:
        \begin{enumerate}
          \item \textbf{P1 (Particle Stress Calculation)}. For each particle, calculate its Piola-Kirchoff stress tensor from its deformation gradient.
          \item \textbf{P2G (Particle to Grid)}. For each particle, interpolate its mass and momentum to the grid with APIC transfer kernels \cite{2015-APIC, jiang2017angular}.
          \item \textbf{G (Grid Velocity Calculation)}. For each grid node, solve for its updated momentum (incorporating any desired external forces) and velocities.
          Any time integration is performed using symplectic Euler.
          \item \textbf{G2P (Grid to Particle)}. For each particle, interpolate its updated velocity and affine matrix from the grid.
          \item \textbf{P2 (Particle Kinematic Update)}. For each particle, update its deformation gradient and position using the updated particle quantities and timestep.
        \end{enumerate}
    We refer the reader to \cite{2018-MLSMPM} and to the appendix for more detailed explanations of the {MLS-MPM} algorithm.

\subsection{Differentiable Simulation} 
    We implemented {MLS-MPM}, which is naturally differentiable; analytical gradients have been derived by Hu et al. \cite{2018-ChainQueen}. 
    We manually implemented the necessary analytical gradients, to have more control over the implementation and avoid potential issues with software-based automatic differentiation \cite{johnson2023software}.
    In our method, these manually derived gradients were then used in conjunction with the non-stochastic version (i.e., all particles are used for each iteration) of the Adam optimizer \cite{kingma2014adam} to perform the optimization, ensuring stable and efficient convergence.
    In addition to optimization, a differentiable simulator naturally requires storing the feedforward simulation network to allow for the backpropagation of derivatives.
    We implemented this network by storing the computation graph of each timestep layer $\textbf{T}^n$, where a layer performs each step of an MPM timestep:

    \begin{equation}
        \textbf{T}^n = \textbf{P1}^n \rightarrow \textbf{P2G}^n \rightarrow \textbf{G}^n \rightarrow \textbf{G2P}^n \rightarrow \textbf{P2}^n .
    \end{equation}
    Backpropagation can then be used on the simulation network to compute gradients via the chain rule:
    \begin{equation}
        \textbf{T}^1 \leftarrow \cdots \leftarrow \textbf{T}^n \leftarrow \mathcal{L} .
    \end{equation}
    We refer to \cite{2018-ChainQueen} for the derivation of the analytical derivatives used in the backward pass. 
    Our method involves controlling particles' deformation gradients (Section \ref{sec:control}), so we introduce a control deformation gradient term separate from the time-evolved deformation gradients.
    The exact modification to the {MPM} algorithm can be seen in the appendix.
    These control deformation gradients are added in a way such that the loss gradient with respect to them is the same as the loss gradient with respect to the time-evolved deformation gradients.
 
\section{Deformation Gradient Control}
\label{sec:control}

    In continuum mechanics, an object has an initial domain in space
    denoted as $\Omega_0$. A point belonging to the object can be defined by its initial position $\textbf{X} \in \Omega_0$. As the object moves through space and time, the world coordinate of a point $\textbf{x}$ at some time $t$ can represented as $\textbf{x} = \phi(\textbf{X}, t) \in \Omega_t$, where $\phi$ is known as a deformation map. The deformation gradient $\textbf{F}$ is defined as the Jacobian of the deformation map:
    \begin{equation}
        \mathbf{F}(\mathbf{X}, t) = \frac{\partial \phi}{\partial \mathbf{X}}(\mathbf{X}, t) .
    \end{equation}
    
    It is common for elastic constitutive models to be defined on
    the deformation gradient $\textbf{F}$ such that internal energy is only built
    up when $\textbf{F}$ represents a non-rigid transformation between the rest
    state $\textbf{X}$ and world state $\textbf{x}$. Our elastic energy function of choice is fixed co-rotational elasticity \cite{2012-FixedCoratedElasticty, jiang2016material}, whose Piola-Kirchoff stress tensor is:
    \begin{equation}
            \textbf{P(F)} = 2\mu(\textbf{F - R}) + \lambda(J-1)J\textbf{F}^{-T} ,
    \end{equation}
    where $J$ is the determinant of $\mathbf{F}$ and $\mathbf{R}$ is its rotational component, which can be computed via the polar decomposition of $\mathbf{F}$ \cite{gast2016implicit}.
    
    Using this elastic model, we can manually control internal forces in our {MPM} simulation through the particle deformation gradients (we are inspired by the preliminary work found in \cite{10.1145/3606037.3606840}).
    An example of this would be manually setting deformation gradients to the identity matrix $\mathbf{F} = \mathbf{I}$ after the shape has deformed. 
    This would correspond to setting the new rest state of the object to the current state. 
    Another way to control internal forces through deformation gradients would be setting expansion ($\alpha > 1$) and contraction ($\alpha < 1$) deformation gradients $\mathbf{F} = \mathbf{\alpha I}$.
    Of course, to actually drive a shape morph, an automatic method for per-particle deformation gradient control is required.
    We achieve this through the optimization of the feedforward simulation network, which requires a loss function to be defined.
    
\subsection{Preliminary Shape Loss Functions}
    To morph a shape from an input to a target geometry, we initially considered two simple loss function choices:

    \textit{Particle positions.} Given a simulation point cloud of size $N$, where
    each material point has a unique index $p \in [1, N]$, we require an
    equally sized target point cloud. A position-based loss function can then be defined as
    the sum of the squared distances between points $\mathbf{x}_p$ in the simulation
    point cloud and points $\mathbf{x}^*_p$ in the target point cloud:
    \begin{equation}
        \mathcal{L} = \sum_p \frac{1}{2}||\mathbf{x}_p - \mathbf{x}_p^*||^2 .
    \end{equation}
    In our experiments, this loss function is suitable for locomotion tasks or simple shape changes such as expansion or shearing;
    however, since it merely penalizes average position errors, using this loss function can leave outlier points in the morph and generally does not provide visually adequate morphs in the presence of fine-scale features or topology changes. 
    Another downside of using the position loss function is the requirement of an equally sized target point cloud with point-to-point correspondences.
    There are techniques available for computing the point-to-point correspondences of two different shapes \cite{2011-ShapeCorrespondenceSurvey}, but this places a constraint on the optimal positions of individual material points.

    \noindent \\ \textit{Nodal mass.} In {MPM}, we have the luxury of a background grid, the nodes of which can store interpolated quantities like mass from the particles.
    Using a nodal mass loss function instead of a point position loss function can alleviate many requirements on our input geometry, such as the need to have identical numbers of particles in the input and target geometries.

    We can then compute the sum of the squared differences between the nodal masses $m_i$ and $m_i^*$ of the interpolated input masses and the interpolated target masses, respectively:
    \begin{equation}
        \mathcal{L} = \sum_{i} \frac{1}{2} (m_i - m^*_i)^2 .
    \end{equation}
 
    This loss function is not easy to optimize with general optimizers and can result in ``mass ejections,'' where clumps of material points are ejected outward.
    We believe that mass ejections are a consequence of converging to the local minima of the nodal mass loss function, where mass ejections lead to the quickest decrease of the loss function.
    This mass ejection problem becomes more noticeable with larger differences between the initial and target shapes.
    In such cases, a stronger force would be applied in a part of the domain with less overlap between input and target geometries (hence, a large difference in mass), resulting in greater mass ejection.

    Knowing this, we designed a small modification to the nodal mass loss function to eliminate mass ejections.

\subsection{Log-based mass loss function} 
     Rather than pursuing various regularization terms to add to the squared error loss, we instead focused on making the loss landscape smoother and less sensitive to outliers.
     We define the log-based mass loss function as 
    \begin{equation}
    \mathcal{L} = \sum_{i} \frac{1}{2} (\ln(m_i + 1) - \ln(m^*_i + 1))^2 .
    \end{equation}
    By transforming grid masses into the $\ln$ of 1 plus the grid masses, where adding $1$ ensures a lower bound for $\ln$ transformation, we successfully eliminate the sensitivity to outliers that led to mass ejection (see our results in Section \ref{sec:examples}).

      \begin{figure*}
            \centering
            \includegraphics[width=\textwidth]{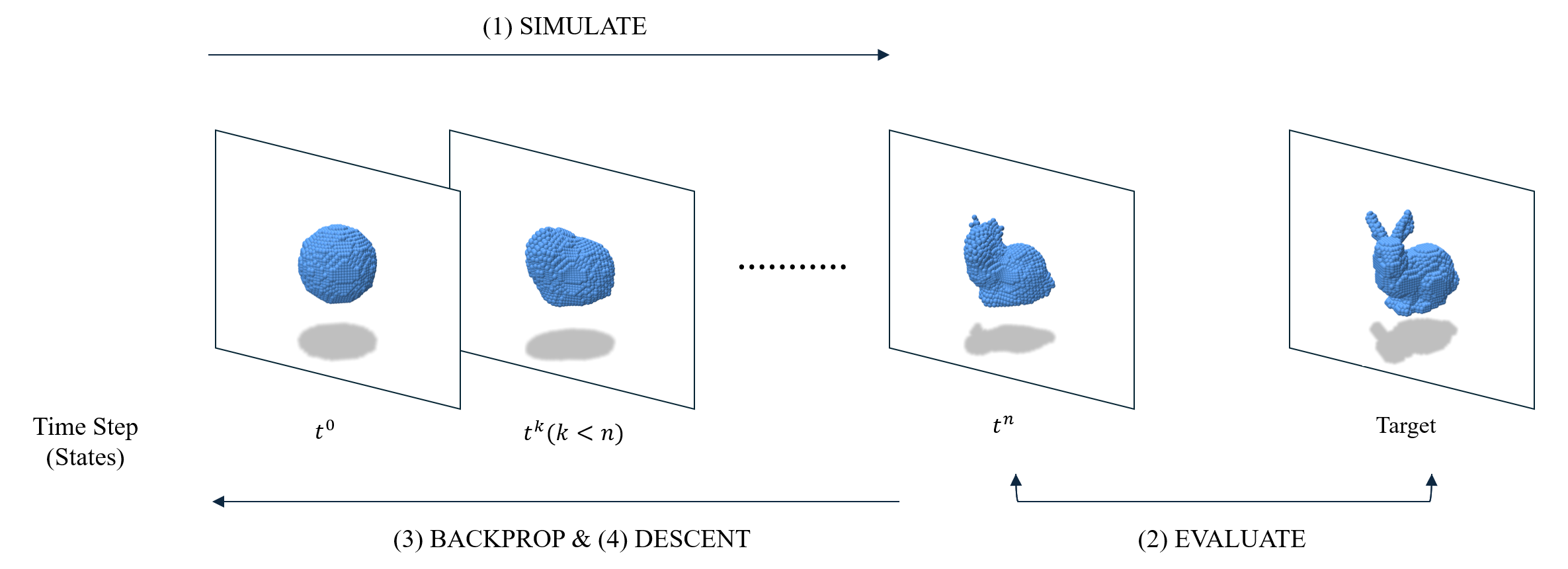}
            \caption{Morphing method. \textmd{Step-by-step optimization process of the simulation networks, aligned according to the causality principle. The Adam optimizer for the control variables is implemented in order for each time step layer of the simulation and includes steps such as forward calculation ({SIMULATE}), loss function calculation ({EVALUATE}), backward calculation ({BACKPROP}), and control variable update ({DESCENT}). This optimization process can be repeated until the norm of the loss gradient falls below a certain tolerance.}}
            \label{fig:Morphing}
        \end{figure*}
        
\section{Morphing Method}

    Given our simulation framework and loss function, we can now describe our proposed morphing algorithm. The main idea is based on Adam optimization of control deformation gradients.
    
\subsection{Simulation Optimization}
    In our simulation network in Figure \ref{fig:Morphing}, the ordering of operations is governed by the principle of causality. This means that any control or intervention applied at earlier stages has a direct impact on the inputs of subsequent layers, and consequently, on their outputs. In this context, ``control'' refers to any manipulation or condition imposed on the system, while ``layers'' represent the sequential stages of the simulation process.
    We outline the steps to perform an iteration of optimization on the control variables for a single timestep layer $\textbf{T}^n$ of the simulation network:
\begin{enumerate}
    \item \textbf{SIMULATE}. Compute the forward pass starting from the control layer to the final layer.
    \item \textbf{EVALUATE}. Compute the loss function via the final layer.
    \item \textbf{BACKPROP}. Compute the backward pass to get the gradient of the loss function with respect to the control variables in the control layer.
    \item \textbf{DESCENT}. Update the control variables by the loss function with Adam optimizer using the origin stepsize $\alpha$, which can be fixed or computed via line search. In our case, we use a bisection line search method (Algorithm \ref{alg:linesearch}) to find a step size that sufficiently decreases loss.
\end{enumerate}
    This process of optimization can be repeated until convergence, which can be defined as the norm of the loss gradient being below a specified tolerance. 
    Once the optimization for a single layer has been satisfactorily completed, the optimization can move on to the next layer.

\subsection{Control Layers}
\label{sec:control-layers}
    To save on computation time, and to allow time for the effects of previously controlled deformation gradients to propagate, we only compute control deformation gradients on a subset of the timesteps. We denote the layers where we compute control deformation gradients as ``control layers.'' For our animations, where we use a simulation frame rate of 120 FPS (and 1 frame = 1 timestep), we chose an interval of 10 timesteps between control layers.

    We outline in pseudocode our algorithm for computing the control deformation gradients using one optimization pass in Algorithm \ref{alg:mpm-morph}.
    To make the notation concise, we let $\tilde{\mathbf{F}}^n$ represent the control deformation gradients at timestep $n$.
    We also let Loss($\tilde{\mathbf{F}}^n$) represent forward simulation from timestep $n$ to $N$, followed by computing the loss between the material points at timestep $N$ and the target shape.
    We note that the partial derivative of the loss is first computed per grid node with respect to the nodal grid mass $m_i$, which will in general be distinct for different grid nodes. 
    Based on this quantity, we can then use MLS-MPM interpolatory weights to compute the derivative of loss with respect to a particle's location $\mathbf{x_p}$, and then apply the chain rule to compute derivatives of the loss with respect to per-particle control deformation gradients, the set of which is $\tilde{\mathbf{F}}^n$.
    The optimization step in Algorithm \ref{alg:mpm-morph} can then update each particle's control deformation gradient. We emphasize that with this algorithm, each particle will, in general, end up with distinct control deformation gradients.

    \begin{algorithm}
        \begin{algorithmic}[1]
        \Require N, the final simulation timestep
        \Require $\Delta n$, the control period
        \Require $i_{\text{max}}$, the max number of iterations per control timestep
        \Require $\kappa$, the convergence tolerance for Adam optimizer
        \Require $\alpha$, the original Adam optimizer step size
        \Require $\beta_1$, $\beta_2$,  the decay rates for the moment estimates
        \Require $\epsilon_{adam}$, a constant to avoid division by zero in Adam
        \State $\tilde{\mathbf{F}}^{1:N} \leftarrow \mathbf{0}$ \Comment{Initialize all control deformation gradients to 0}
        \State $m^{1:N} \leftarrow \mathbf{0}$ \Comment{Initialize first moment vector}
        \State $v^{1:N} \leftarrow \mathbf{0}$ \Comment{Initialize second moment vector}
        \State $t \leftarrow 0$
        \State $n \leftarrow 1$
        \While{$n < N$}
            \While{$i < i_{\text{max}}$}
                \State $L \leftarrow \text{Loss}(\tilde{\mathbf{F}}^n)$ \Comment{Calculate the loss after simulation from $n$ to $N$ using $\tilde{\mathbf{F}}^n$}
                \State $\mathbf{g} \leftarrow \frac{\partial L}{\partial \tilde{\mathbf{F}}^n}$ \Comment{Compute gradients}
                
                \State $m^n \leftarrow \beta_1 \cdot m^n + (1 - \beta_1) \cdot \mathbf{g}$ 
                \State $v^n \leftarrow \beta_2 \cdot v^n + (1 - \beta_2) \cdot (\mathbf{g} \odot{} \mathbf{g})$ \Comment{Hadamard product (elementwise multiplication)}
                
                \State $\hat{m}^n \leftarrow \frac{m^n}{1 - \beta_1^t}$ 
                \State $\hat{v}^n \leftarrow \frac{v^n}{1 - \beta_2^t}$ 
                
                \State $\alpha \leftarrow \text{ComputeStepSize}(L, \tilde{\mathbf{F}}^n, \hat{m}^n, \alpha)$
        
                \State $\tilde{\mathbf{F}}^n \leftarrow \tilde{\mathbf{F}}^n - \alpha \frac{\hat{m}^n}{\sqrt{\hat{v}^n} + \epsilon_{adam}}$ \Comment{Update $\tilde{\mathbf{F}}^n$}
        
                \If{$||\mathbf{g}|| < \kappa$}
                    \State \textbf{break}
                \EndIf
                \State $i \leftarrow i + 1$ \Comment{Proceed to the next Adam iteration}
            \EndWhile
            \State $n \leftarrow n + \Delta n$ \Comment{Advance to the next control layer}
         \EndWhile
        \State \textbf{return} $\tilde{\mathbf{F}}^{1:N}$ \Comment{Final optimized $\tilde{\mathbf{F}}^n$ for all timesteps}
        \end{algorithmic}
        \caption{Control Deformation Gradient Computation}\label{alg:mpm-morph}
    \end{algorithm}

    \begin{algorithm}
    \begin{algorithmic}[1]
        \Require $L_0$, the original loss
        \Require $\mathbf{x}$, the original vector argument
        \Require $\Delta \mathbf{x}$, the step direction vector
        \Require $\alpha$, the original step size
        \State $L \leftarrow $ Loss$(\mathbf{x} + \alpha \Delta \mathbf{x})$
        \While{$L > L_0$}
            \State $\alpha \leftarrow \alpha / 2$
            \State $L \leftarrow \text{Loss}(\mathbf{x} + \alpha \Delta \mathbf{x})$
        \EndWhile
        \State \textbf{return} $\alpha$ 
        \end{algorithmic}
    \caption{ComputeStepSize ($L_0$, $\mathbf{x}$, $\Delta \mathbf{x}$, $\alpha$)}\label{alg:linesearch}
    \end{algorithm}

\subsection{Multiple Optimization Passes}
    As described in the prior subsection, we use the routine of Algorithm \ref{alg:mpm-morph} to compute optimal control deformation gradients for each control layer.
    However, this strategy is suboptimal: for instance, when we compute optimal control deformation gradients for the first control layer, we are using values of 0 for the control deformation gradients at all subsequent control layers, when we know that we will later optimize those to also have non-zero values.
    Accordingly, to obtain improved control deformation gradient values, we can utilize a multi-pass optimization scheme where, after optimizing control deformation gradients at all control layers, we can go through and re-run the optimization routine of Algorithm \ref{alg:mpm-morph} using the computed $\tilde{\mathbf{F}}^{1:N}$ as initial/improved guesses for the optimization.
    
    For a clear analysis of the tradeoffs, we evaluated the impact of the number of passes and the number of gradient iterations on performance based on the data shown in Table \ref{tab:multipass}. This ablation study provides useful insights into the relationship between loss values and execution time. For example, there is a significant improvement in loss between 1 pass and 3 passes, while the execution time remains the same. This suggests that multi-pass optimization can improve the loss without degrading performance.

    Additionally, our analysis of the limitations of increasing the number of passes shows that beyond 6 or 12 passes, the loss values do not improve and may even increase. This indicates the limits of optimization, highlighting that a larger number of passes does not always guarantee better results.
    
    Lastly, we observed that despite varying the number of passes and gradient iterations, the execution time remains nearly constant. This is a significant finding, as it emphasizes that further optimization can reduce the loss without negatively impacting performance.
    
    \begin{table}[ht!]
        \centering
        \small
        \caption{Varying the number of optimization passes and optimization iterations per pass.}
        \begin{tabular}{cccc}
        \toprule
        Passes & \# of Gradient Iter & Final Loss & Elapsed Time [s] \\
        \midrule
        1   & 12 & 92.3826 & 187 \\
        3   & 4  & 91.0895 & 187 \\
        6   & 2  & 93.4914 & 187 \\
        12  & 1  & 93.1483 & 186 \\
        \bottomrule
        \end{tabular}
        \label{tab:multipass}
    \end{table}
    
\subsection{Chaining Multiple Simulation Networks}
    The amount of memory required for a simulation network is directly proportional to the number of simulation timesteps.
    This can make long animations require an infeasible amount of memory.
    Furthermore, backpropagation through longer networks can result in accumulating floating point errors that lead to problems like vanishing gradients \cite{hochreiter1998vanishing}. 

    Therefore, for longer morph animations, we can divide the morph into different segments and optimize one segment at a time.
    We construct a new simulation network using the final layer of the previous optimized network.
    The animation produced by the previous network is saved to disk before the memory of the previous network is freed.
    The new network is then optimized, and this process repeats.
    In our experiments, we split the optimization for long animations into networks of 10 timesteps each, i.e., 0.083-second segments at 120 FPS.
    By chaining optimized simulation networks together, we can produce indefinitely longer animations without any increased memory requirements and with reduced computational cost.

\section{Examples}

\begin{table*}[!ht]
    \centering
        \scriptsize
        \caption{Parameters for the various examples shown in Figure \ref{fig:Overall_deform}. Density and smoothing factor $\gamma$ varied slightly between examples to improve the smoothness of the morph.}
        \begin{tabularx}{1.0\textwidth}{@{}lcccccccc@{}}
            \toprule
            Example & \# of Frames & Density [kg/$m^3$] & $\gamma$ & dt &\# of Gradient Iter (\# of Path) & Target \# of Particles & Initial \# of Particles & Grid Resolution \\
            \midrule
            Sphere to Bunny	& 420 & 75 & 0.955 & 0.00833 & 4(3) & $1.116 \times 10^4 $ & $0.92 \times 10^4 $ & $32 \times 32 \times 32$ \\ 
            Sphere to Bob   & 600 & 60 & 0.955 & 0.00833 & 4(3) & $1.116 \times 10^4 $ & $1.06 \times 10^4$ & $32 \times 32 \times 32$ 	\\ 
            Sphere to Spot  & 400 & 60 & 0.955 & 0.00833 & 4(3) & $1.116 \times 10^4$ & $0.76 \times 10^4$ & $32 \times 32 \times 32$ 	\\ 
            Sphere to Car   & 450  & 60 & 0.955  & 0.00833 & 4(3) & $1.116 \times 10^4$ & $1.21 \times 10^4$ & $32 \times 32 \times 32$ 	\\ 
            
            \midrule
            Bob to Sphere   & 450 & 70 & 0.950 & 0.00833 & 4(3) & $1.06 \times 10^4$ & $1.11 \times 10^4$ & $32 \times 32 \times 32$ 	  \\ 
            Bob to Bunny    & 600 & 75 & 0.955 & 0.00833 & 4(3) & $1.06 \times 10^4$ & $0.92 \times 10^4 $ & $32 \times 32 \times 32$       \\ 
            Bob to Spot	    & 500 & 75	& 0.955  & 0.00833 & 4(3) & $1.06 \times 10^4$ & $0.76 \times 10^4$ & $32 \times 32 \times 32$ 	    \\ 
            Bob to Car	    & 600 & 60	& 0.940  & 0.00833 & 4(3) & $1.06 \times 10^4$ & $1.06 \times 10^4$ & $32 \times 32 \times 32$ 	    \\ 
            
            \midrule
            Bunny to Sphere	& 500 & 100 & 0.965 & 0.00833 & 4(3) & $0.92 \times 10^4 $ & $1.116 \times 10^4$ & $32 \times 32 \times 32$    \\ 
            Bunny to Bob	& 630 & 120 & 0.96 & 0.00833 & 4(3) & $0.92 \times 10^4 $ & $1.116 \times 10^4$ & $32 \times 32 \times 32$ 	\\ 
            Bunny to Spot	& 400 & 110 & 0.955 & 0.00833 & 4(3) & $0.92 \times 10^4 $ & $0.76 \times 10^4 $ & $32 \times 32 \times 32$ \\ 
            Bunny to Car	& 850 & 75 & 0.955 & 0.00833 & 4(3) & $0.92 \times 10^4 $ & $1.21 \times 10^4$ & $32 \times 32 \times 32$ 	\\ 
            
            \midrule			
            Spot to Sphere	& 400	& 75 & 0.955 & 0.00833 & 4(3) & $0.76 \times 10^4$ & $1.116 \times 10^4$ & $32 \times 32 \times 32$  \\ 
            Spot to Bunny	& 400	& 75 & 0.955 & 0.00833 & 4(3) & $0.76 \times 10^4$ & $0.92 \times 10^4 $ & $32 \times 32 \times 32$  \\ 
            Spot to Bob	    & 600	& 75 & 0.955 & 0.00833 & 4(3) & $0.76 \times 10^4$ & $1.06 \times 10^4$ & $32 \times 32 \times 32$ 	 \\ 
            Spot to Car	    & 600	& 75 & 0.955 & 0.00833 & 4(3) & $0.76 \times 10^4$ & $1.21 \times 10^4$ & $32 \times 32 \times 32$ 	 \\ 
            \midrule			
            Car to Sphere	& 600   & 75 & 0.950  & 0.00833 & 4(3) & $1.21 \times 10^4$ & $1.116 \times 10^4$ & $32 \times 32 \times 32$      \\ 
            Car to Bunny	& 600	& 75 & 0.930  & 0.00833 & 4(3) & $1.21 \times 10^4$ & $0.92 \times 10^4$ & $32 \times 32 \times 32$  	\\ 
            Car to Bob	    & 300	& 60 & 0.955  & 0.00833 & 4(3) & $1.21 \times 10^4$ & $1.06 \times 10^4$ & $32 \times 32 \times 32$    \\ 
            Car to Spot	    & 600	& 75 & 0.955 & 0.00833 & 4(3) & $1.21 \times 10^4$ & $0.76 \times 10^4$ & $32 \times 32 \times 32$   \\ 
            \bottomrule
        \end{tabularx}
    \vspace{1mm}
    \label{tab:params}
\end{table*}

\label{sec:examples}

    We demonstrate our methodology on a wide range of physically-based morphing animations in three-dimensional space.
    Leveraging MPM, our objects can seamlessly undergo topological morphs driven by internal elastic forces.
    We use a modified explicit time integration scheme for {MLS-MPM} (see the appendix), which helps prevent instability when taking larger timesteps.
    A full list of variables involved in our {MPM} implementation is also found in the appendix.

    Parameters for all the experiments shown in Figure \ref{fig:Overall_deform} are summarized in Table \ref{tab:params}.
    The table lists the number of particles in both the initial and target meshes.
    While the number of particles does not change during the morphing process---the particle count used for each animation is fixed to the number of particles in the initial mesh---the number of particles in the target mesh can also affect performance (so we report this number as well).
    This is because the target mesh particles are used in computations of the various derivatives of the loss function used during the backpropagation step of our algorithm.
    For all our examples, MPM point cloud visualization was performed using Polyscope \cite{polyscope}, and rendering was conducted using Houdini.
    We ran all experiments on a machine with an Intel i9-13950HX CPU, 32GB RAM, and an NVIDIA RTX 4090 GPU.

\begin{figure}
        \centering
        \includegraphics[width=\linewidth]{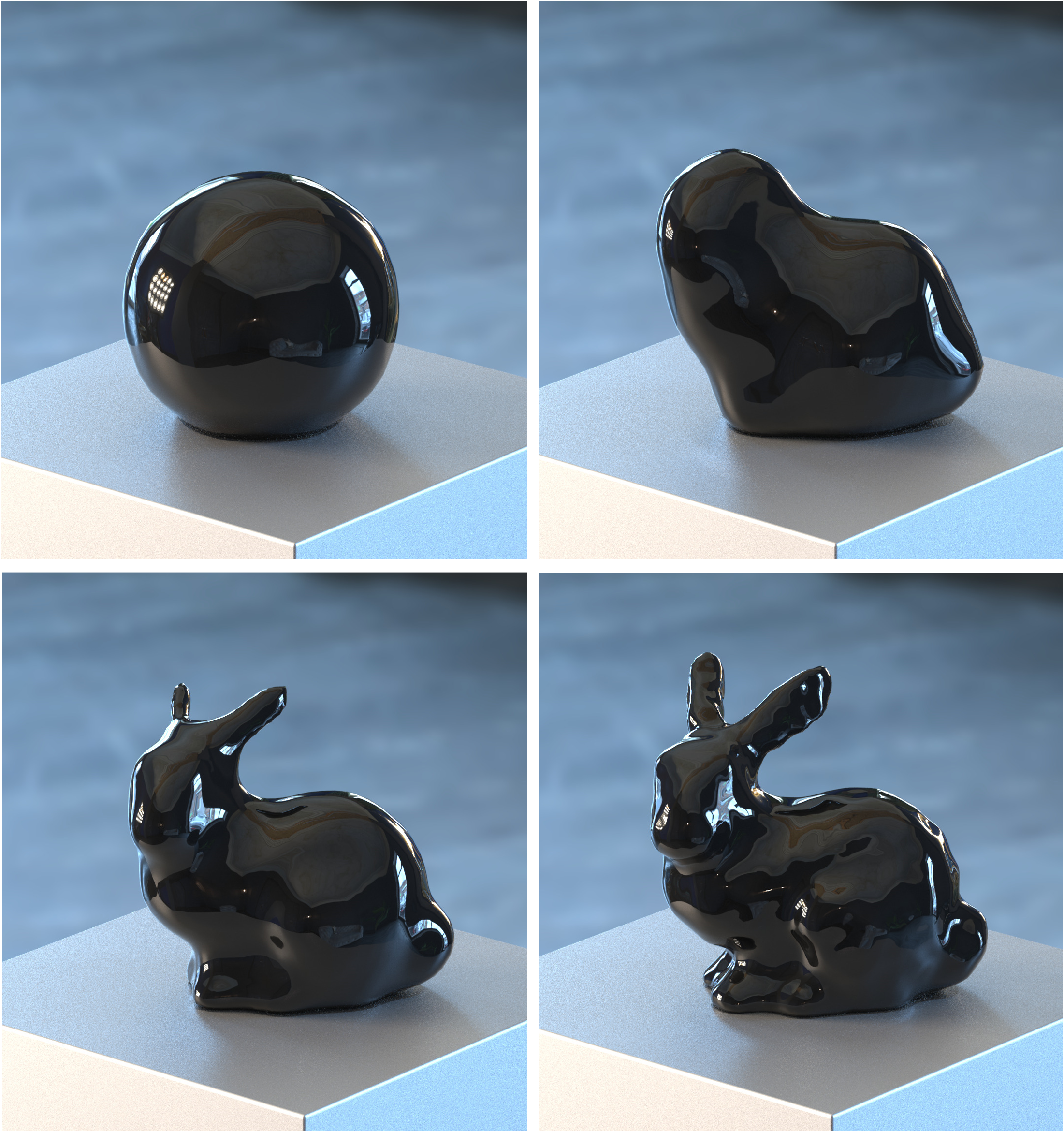}
        \caption{Sphere to bunny morphing animation. Key stages show the sphere evolving into a detailed bunny form, highlighting our method's precision in capturing sharp features like the ears.}
        \label{fig:Sphere@Bunny}
    \end{figure}

    \textit{Sphere to Bunny.} In our first example,  we present a morphing animation from a sphere to a bunny, as depicted in Figure \ref{fig:Sphere@Bunny}. This demonstrates a successful shape morph using our log-based nodal mass loss function (see supplemental video). While using other loss functions might result in phenomena like mass ejection, our method, as shown in the figure, allows for smooth transformation without ejected particles or other unpleasant artifacts.

    One of the most notable aspects of this example is the representation of the bunny's ears.
    Our methodology allows us not only to match the bulk volume of a target mesh but also to faithfully reproduce its sharp and thin features.

    \begin{figure}
        \centering
        \includegraphics[width=\linewidth]{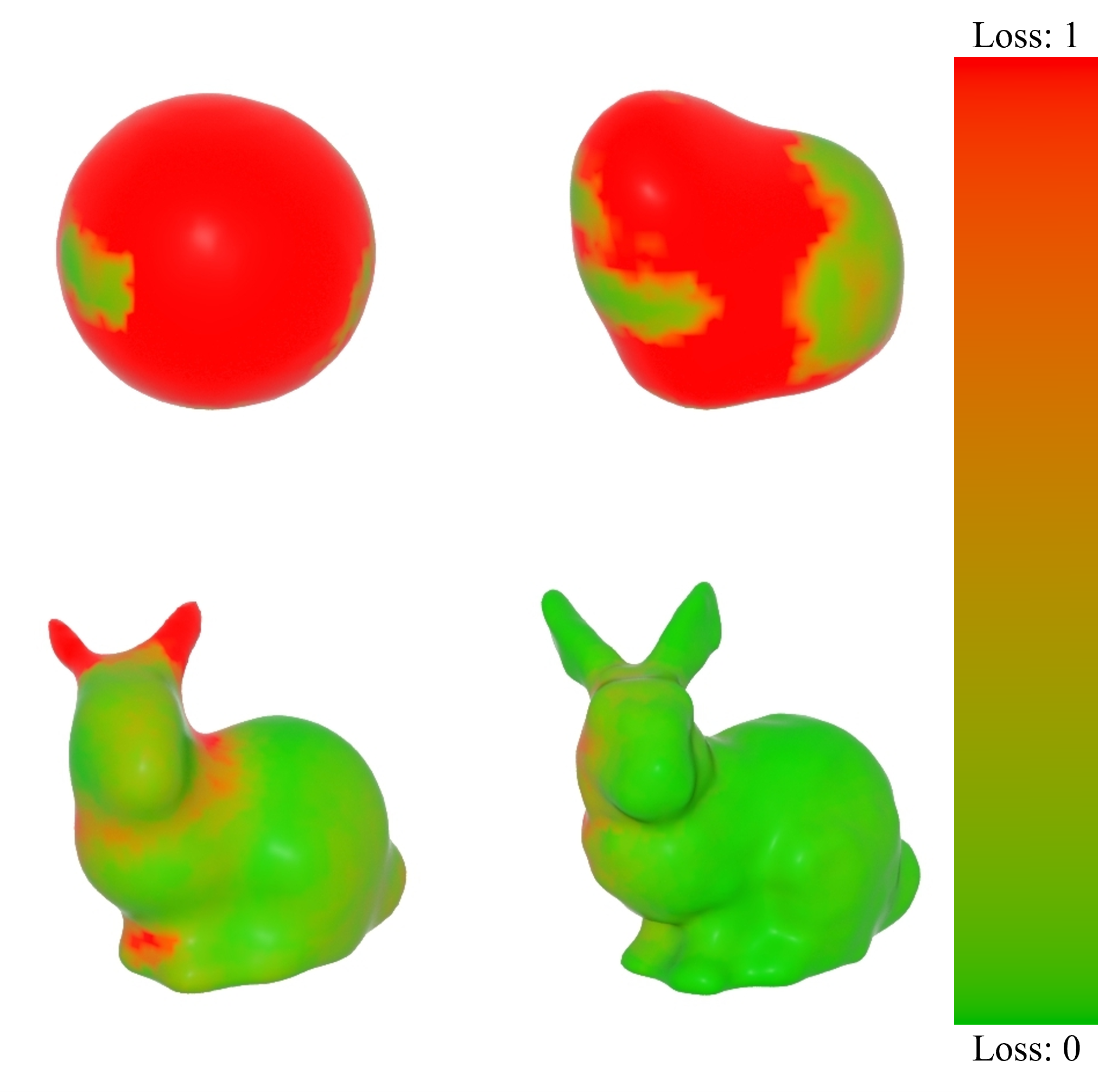}
        \caption{Visualizing our loss function while morphing a sphere to a bunny. Loss is normalized from 0 to 1 for ease of visualization. We observe a smooth decrease in the loss, including in the finer regions of the target geometry towards the end of the morph.}
        \label{fig:Loss}
    \end{figure}

    We also plot the loss function, interpolated to the surface, as the geometry deforms; see Figure \ref{fig:Loss}.
    The loss function is normalized from 0 to 1 in this figure for ease of visualization.
    We observe that the loss function smoothly decreases as the morph proceeds, with fine-scale details of the target mesh generally being the last parts of the morph to reach an acceptable state.
    
\begin{figure}
        \centering
        \includegraphics[width=\linewidth]{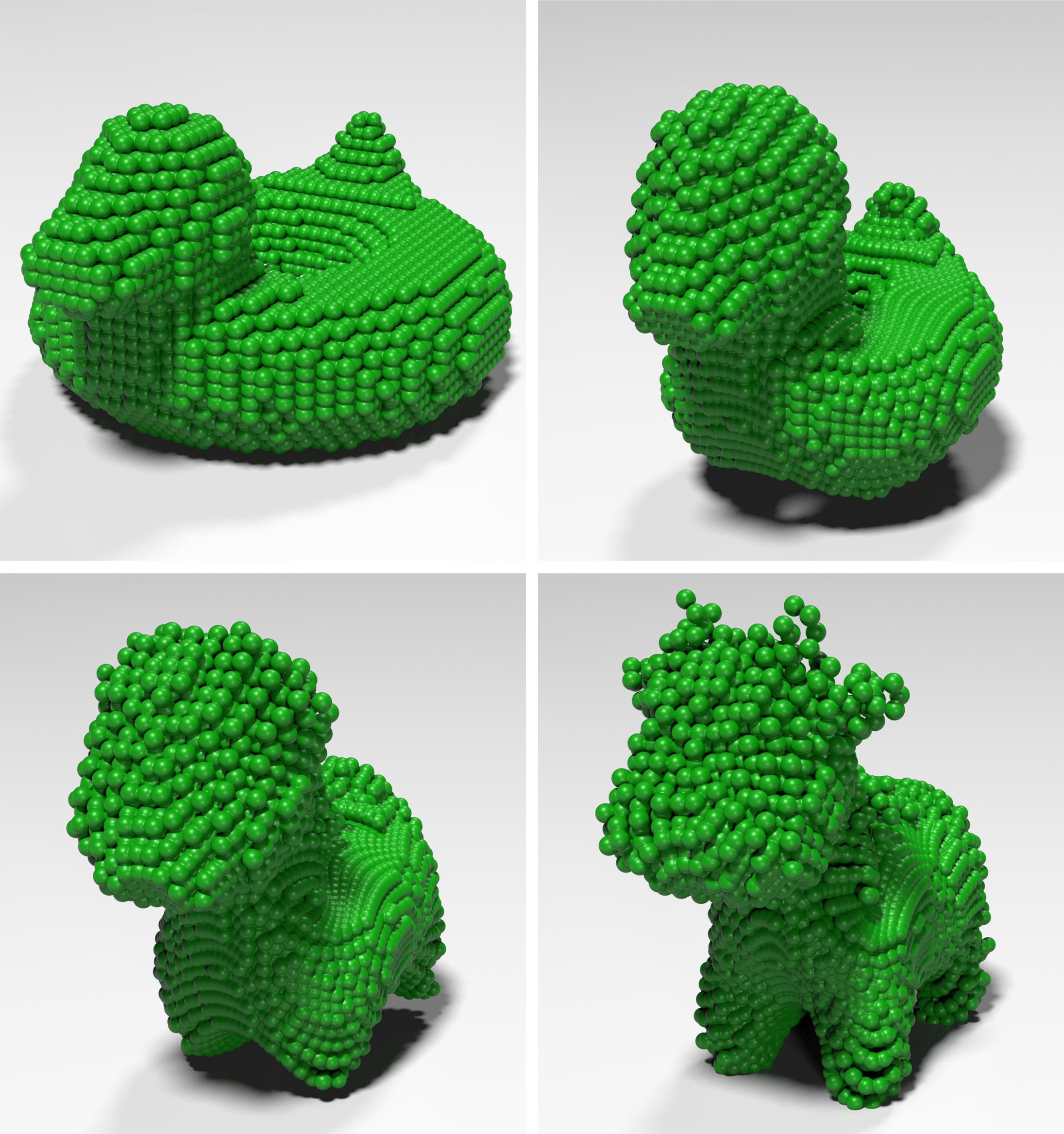}
        \caption{Morphing Duck (Bob) to Cow (Spot). \textmd{As the model deforms into a cow, we observe a smooth topology change as the center hole of the inflatable duck closes. This example evidences our method's ability to handle simultaneously evolving geometry and topology.}}
        \label{fig:Duck@Cow}
    \end{figure}

    \textit{Duck (Bob) to Cow (Spot).} In the second example, we illustrate an animation transitioning from a duck model (``Bob'') to a cow model (``Spot'') in Figure \ref{fig:Duck@Cow}.
    This case is interesting due to the changing topology between the two geometries.
    Our method achieves a smooth morph of the geometry and topology simultaneously, which is a natural benefit of our underlying {MPM} framework.

    \begin{table}[ht!]
      \scriptsize
      \caption{Performance for various shape morphing sequences, including accuracy.}
      \begin{tabular}{lcccc}
       \toprule
       Example & Elapsed Times [s] & Initial Loss & Final Loss & Acc. [\%] \\
       \midrule
        Sphere to Bunny&    187  &	5302.69  &	91.0895  &	98.28 \\ 
        Sphere to Bob&	    290  &	11600   &	300.528  &	97.41 \\ 
        Sphere to Spot&	    172  &	7685.97  &	124.435  &	98.38 \\ 
        Sphere to Car&	    188  &	8163.73  &	104.355  &	98.72 \\ 
        \midrule	   	
        
        Bob to Sphere&	    173  &	12018.7  &	388.199  &	96.77 \\ 
        Bob to Bunny&	    332  &	12055.2  &	166.26   &	98.62 \\ 
        Bob to Spot&	    195  &	12218.3  &	206.332  &	98.31 \\ 
        Bob to Car&	        266  &	11516.7  &	368.944  &	96.80 \\ 
        \midrule			
        
        Bunny to Sphere&	223  &	6104.12  &	119.751  &	98.04 \\ 
        Bunny to Bob&	    278  &	14184.2  &   608.073  &	95.71 \\ 
        Bunny to Spot&	    165  &	4469.88  &	70.51    &	98.42 \\ 
        Bunny to Car&	    415  &	11945.2  &	870.81   &	92.71 \\ 
        \midrule		    	
        Spot to Sphere&	    123  &	8606.47  &  465.712  &	94.59 \\ 
        Spot to Bunny&	    120  &	3694.31  &	116.349  &	96.85 \\ 
        Spot to Bob&	    178  &	12218.3  &	575.581  &	95.29 \\ 
        Spot to Car&        204  &	14316.9  &	685.355  &	95.21 \\ 
        \midrule
        Car to Sphere&	    333  &	9980.66  &	122.113  &	98.78 \\ 
        Car to Bunny&	    335  &	11945.2  &	108.534  &	99.09 \\ 
        Car to Bob&	        309  &	11516.7  &	398.1    &	96.54 \\ 
        Car to Spot&	    333  &	14316.9  &	167.973  &	98.83 \\ 
         \bottomrule
      \end{tabular}
      \vspace{1mm}
      \label{tab:performance}
    \end{table}

    \textit{Morphing between various mesh pairs; smoothing factor.} To further test our method, we explored pairwise deformations across a diverse set of five distinct meshes; see Figure \ref{fig:Overall_deform}.
    We aimed to rigorously assess the efficacy and reliability of our computational framework under varying geometric conditions.
    All attempted morphs were successful, demonstrating the robustness and adaptability of our approach.
    An additional element in achieving such a high degree of success was the interpolation of a deformation gradient mechanism ($\textbf{F}^{n+1} = (1 - \gamma)\textbf{F}^{n+1} + \gamma\textbf{F}^{n}$) during the post-processing phase. 
    
    The smoothness values $\gamma$ we chose are reported in Table \ref{tab:params}, and were hand-tuned per example (generally keeping them as small as possible, while ensuring a smooth morphing animation).

    \textit{Performance and parallelization.} Another benefit of our method is that the majority of our algorithm is parallelizable, due to our foundation on {MPM}. We produced a CPU-parallel implementation using OpenMP, leading to significantly shorter runtimes than a serial code.
 
\begin{figure}[ht!]
        \centering
        \includegraphics[width=\linewidth]{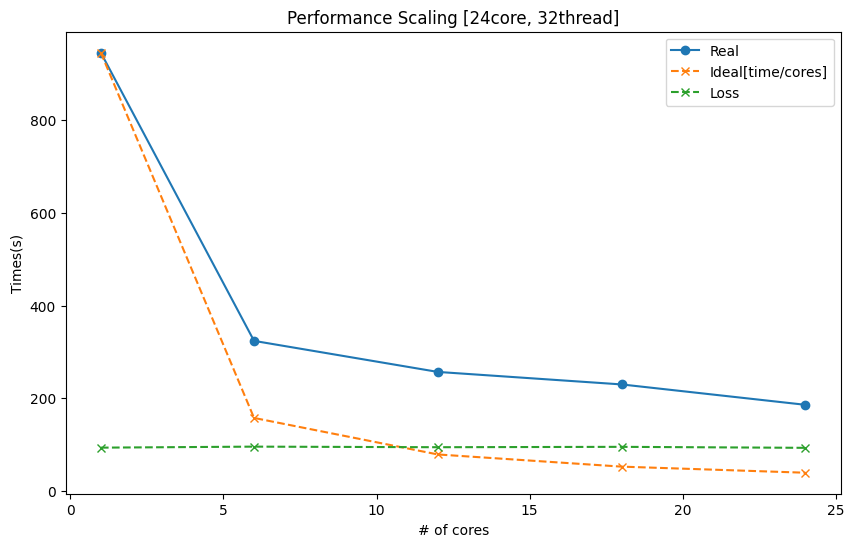}
        \caption{Parallel scaling. \textmd{As we run an example of our method with more cores, we observe nearly linear scaling.  The dashed green line plots the difference between actual time (blue, solid line) and theoretical time under perfect linear scaling (yellow dashed line).  Our method is not fully parallelized, so we do not expect linear scaling.}}
        \label{fig:Core_per_time}
    \end{figure}

    The performance of our parallel code on the various examples is reported in Table \ref{tab:performance}. 
    We also demonstrate the strong scaling of our method in Figure \ref{fig:Core_per_time}; while our algorithm is not fully parallelizable and hence we do not expect perfect linear scaling, we still observe useful scaling from 1 to 24 cores (over 5x speedup).

    \textit{Complex Object Animation.} In our complex object animation examples, we demonstrate that our approach can generate smooth morphing animations for complex surfaces and dynamic topology scenarios. In these examples, we test both letter models and a dragon model, which possess curves and topologies that could pose challenges for traditional shape morphing methods. The letter models showcase our method's ability to handle sharp edges and changing topologies, while the dragon model tests its capability to preserve intricate details during the morphing process.

\begin{figure}[!bt]
        \centering
        \includegraphics[width=\linewidth]{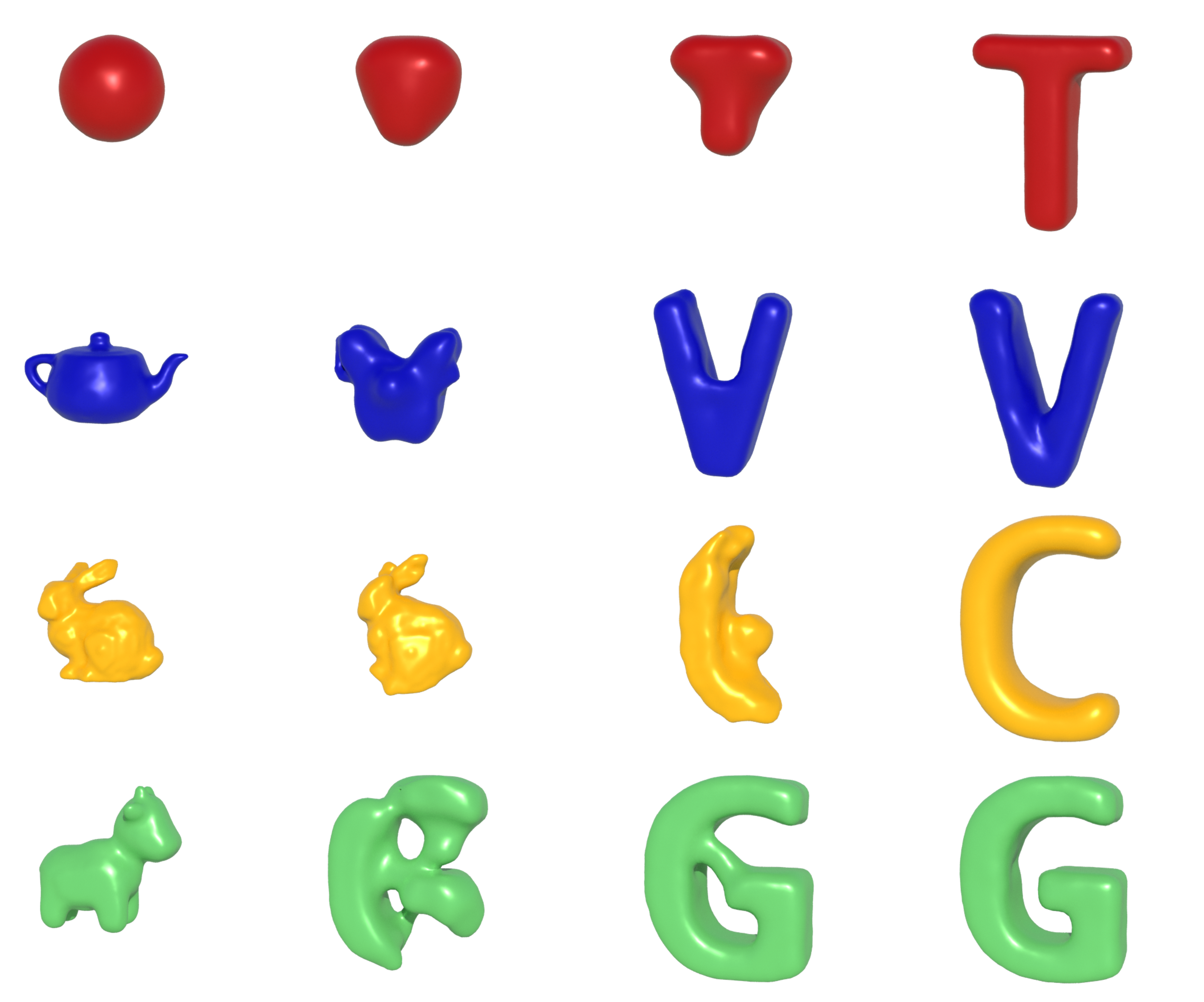}
        \caption{Morphing example with letters. This example demonstrates the capability of our morphing technique to capture topological changes and fine details during a morph. The sequence maintains accurate detail as each form transitions into the next, showing the potential for making practical animations with our method.}
        \label{fig:TVCG}
    \end{figure}

\begin{figure}[!bt]
        \centering
        \includegraphics[width=\linewidth]{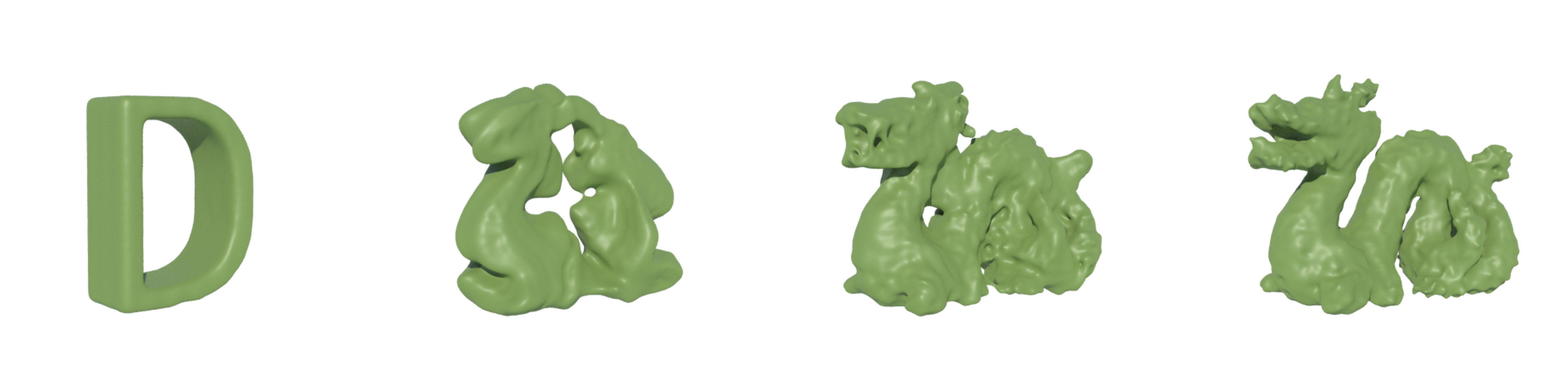}
        \caption{Morphing dragon example. This example demonstrates the transformation from the letter `D' into a dragon model. Our method effectively handles the complex topology changes that occur during the morph, preserving both surface details and structural coherence.}
        \label{fig:Dragon}
    \end{figure}

    Our approach successfully generates these animations (see Figures \ref{fig:TVCG} and \ref{fig:Dragon}).
    For the dragon model, we utilized a larger number of particles to better capture details throughout the morphing process (see Table \ref{tab:numofparticles}).
    We rendered this animation using the VDB from Particles node in Houdini with a varying voxel size and applied ten iterations of Laplace smoothing to the resulting mesh on each frame for better visual fidelity.
    
   \begin{table}[!bt]
      \scriptsize
      \centering
      \caption{Number of Particles for our complex examples}
      \begin{tabular}{lccc}
       \toprule
       Example & \# of Frames & Initial \# of Particles & Target \# of Particles\\
       \midrule
        Sphere to T&    600  &	$8.91 \times 10^4$  &	$9.31 \times 10^4$ \\ 
        Teapot to V&	900  &	$7.06 \times 10^4$  &	$9.23 \times 10^4$ \\ 
        Bunny to C&	    300  &	$5.42 \times 10^4$  &	$8.26 \times 10^4$ \\ 
        Spot to G&	    1000  &	$6.12 \times 10^4$  &	$14.0 \times 10^4$  \\ 
        D to Dragon & 1000    & $2.46 \times 10^4$  &   $3.92 \times 10^4$\\
        \bottomrule
      \end{tabular}
      \vspace{1mm}
      \label{tab:numofparticles}
    \end{table}

\section{Conclusions, Limitations, and Future Work}

    This study presented a framework for shape morphing based on deformation gradient control in the differential material point method. Our method's advantages include the natural handling of topological changes during a morph, as well as the ability to match fine-scale details of the target mesh.
    Our method also inherits {MPM}'s parallelization properties, enabling more efficient implementations.
    
    Key to our method is the novel log-based nodal mass loss function we used, which prevents spurious particle movement due to relatively large distances between pieces of an input and target geometry. Our control layer and chained optimization approaches enable fine-grained control of the morph and the creation of long-duration morphs within our framework.
    
    Techniques like gradient smoothing factor also help ensure smooth morphing animations. To further enhance the accuracy of the optimization, we applied the smoothing factor not only during forward passes but also in back-propagation, ensuring that the optimization process benefits from the same smoothing during gradient updates.

    We are interested in exploring different regularizing techniques, such as a different loss function that incorporates deformation gradient information in a helpful way.
    We also note that while our method scaled reasonably well up to about 24 cores with OpenMP, in order to better serve real artist use cases, it is important to improve the scaling of our method up to a few dozen cores.  We are interested in more carefully analyzing the parallel performance and bottlenecks of our algorithm, as well as producing a CUDA-based implementation that leverages GPUs.
    Works like \cite{wang2020massively} suggest that a well-tuned GPU {MPM} implementation should enable orders-of-magnitude speedups over the timings reported in the present work, even with no substantial algorithmic changes.
   
    Shape morphing remains a practical task in computer animation, and {MPM} continues to be a popular technique in graphics (e.g., animating the snow in \textit{Frozen} (2013)).
    We anticipate that continued investigation of {MPM}-based shape morphing algorithms will lead to further algorithmic improvements and practical animation tools.

\bibliographystyle{IEEEtran}
\bibliography{ref}

\begin{thebibliography}{10}
\providecommand{\url}[1]{#1}
\csname url@samestyle\endcsname
\providecommand{\newblock}{\relax}
\providecommand{\bibinfo}[2]{#2}
\providecommand{\BIBentrySTDinterwordspacing}{\spaceskip=0pt\relax}
\providecommand{\BIBentryALTinterwordstretchfactor}{4}
\providecommand{\BIBentryALTinterwordspacing}{\spaceskip=\fontdimen2\font plus
\BIBentryALTinterwordstretchfactor\fontdimen3\font minus
  \fontdimen4\font\relax}
\providecommand{\BIBforeignlanguage}[2]{{%
\expandafter\ifx\csname l@#1\endcsname\relax
\typeout{** WARNING: IEEEtran.bst: No hyphenation pattern has been}%
\typeout{** loaded for the language `#1'. Using the pattern for}%
\typeout{** the default language instead.}%
\else
\language=\csname l@#1\endcsname
\fi
#2}}
\providecommand{\BIBdecl}{\relax}
\BIBdecl

\bibitem{gomes1999warping}
J.~Gomes, L.~Darsa, B.~Costa, and L.~Velho, \emph{{Warping and morphing of
  graphical objects}}.\hskip 1em plus 0.5em minus 0.4em\relax San Francisco,
  CA, USA: {Morgan Kaufmann Publishers Inc.}, 1998.

\bibitem{doi:10.1137/22M1524254}
\BIBentryALTinterwordspacing
G.~Friesecke and M.~Penka, ``{The GenCol Algorithm for High-Dimensional Optimal
  Transport: General Formulation and Application to Barycenters and Wasserstein
  Splines},'' \emph{SIAM Journal on Mathematics of Data Science}, vol.~5,
  no.~4, pp. 899--919, 2023. [Online]. Available:
  \url{https://doi.org/10.1137/22M1524254}
\BIBentrySTDinterwordspacing

\bibitem{10.1145/2766963}
\BIBentryALTinterwordspacing
J.~Solomon, F.~de~Goes, G.~Peyr\'{e}, M.~Cuturi, A.~Butscher, A.~Nguyen, T.~Du,
  and L.~Guibas, ``{Convolutional Wasserstein Distances: Efficient Optimal
  Transportation on Geometric Domains},'' \emph{ACM Trans. Graph.}, vol.~34,
  no.~4, jul 2015. [Online]. Available: \url{https://doi.org/10.1145/2766963}
\BIBentrySTDinterwordspacing

\bibitem{terzopoulos1988deformable}
\BIBentryALTinterwordspacing
D.~Terzopoulos and K.~Fleischer, ``{Deformable models},'' \emph{{The Visual
  Computer}}, vol.~4, no.~6, pp. 306--331, 1988. [Online]. Available:
  \url{https://doi.org/10.1007/BF01908877}
\BIBentrySTDinterwordspacing

\bibitem{treuille2003keyframe}
\BIBentryALTinterwordspacing
A.~Treuille, A.~McNamara, Z.~Popovi\'{c}, and J.~Stam, ``{Keyframe Control of
  Smoke Simulations},'' \emph{ACM Trans. Graph.}, vol.~22, no.~3, p. 716–723,
  jul 2003. [Online]. Available: \url{https://doi.org/10.1145/882262.882337}
\BIBentrySTDinterwordspacing

\bibitem{2004-FluidControlAdjointMethod}
\BIBentryALTinterwordspacing
A.~McNamara, A.~Treuille, Z.~Popovi\'{c}, and J.~Stam, ``{F}luid {C}ontrol
  {U}sing the {A}djoint {M}ethod,'' \emph{ACM Trans. Graph.}, vol.~23, no.~3,
  p. 449–456, 8 2004. [Online]. Available:
  \url{https://doi.org/10.1145/1015706.1015744}
\BIBentrySTDinterwordspacing

\bibitem{stomakhin2017fluxed}
\BIBentryALTinterwordspacing
A.~Stomakhin and A.~Selle, ``{Fluxed Animated Boundary Method},'' \emph{ACM
  Trans. Graph.}, vol.~36, no.~4, jul 2017. [Online]. Available:
  \url{https://doi.org/10.1145/3072959.3073597}
\BIBentrySTDinterwordspacing

\bibitem{2012-DeformableObjectsAlive}
\BIBentryALTinterwordspacing
S.~Coros, S.~Martin, B.~Thomaszewski, C.~Schumacher, R.~Sumner, and M.~Gross,
  ``{D}eformable {O}bjects {A}live!'' \emph{ACM Trans. Graph.}, vol.~31, no.~4,
  7 2012. [Online]. Available: \url{https://doi.org/10.1145/2185520.2185565}
\BIBentrySTDinterwordspacing

\bibitem{min2019softcon}
\BIBentryALTinterwordspacing
S.~Min, J.~Won, S.~Lee, J.~Park, and J.~Lee, ``{SoftCon: Simulation and Control
  of Soft-Bodied Animals with Biomimetic Actuators},'' \emph{ACM Trans.
  Graph.}, vol.~38, no.~6, nov 2019. [Online]. Available:
  \url{https://doi.org/10.1145/3355089.3356497}
\BIBentrySTDinterwordspacing

\bibitem{raveendran2012controlling}
\BIBentryALTinterwordspacing
K.~Raveendran, N.~Thuerey, C.~Wojtan, and G.~Turk, ``{Controlling Liquids Using
  Meshes},'' in \emph{Eurographics/ ACM SIGGRAPH Symposium on Computer
  Animation}, J.~Lee and P.~Kry, Eds.\hskip 1em plus 0.5em minus 0.4em\relax
  The Eurographics Association, 2012. [Online]. Available:
  \url{http://dx.doi.org/10.2312/SCA/SCA12/255-264}
\BIBentrySTDinterwordspacing

\bibitem{turk2005shape}
\BIBentryALTinterwordspacing
G.~Turk and J.~F. O'Brien, ``{Shape Transformation Using Variational Implicit
  Functions},'' in \emph{Proceedings of the 26th Annual Conference on Computer
  Graphics and Interactive Techniques}, ser. SIGGRAPH '99.\hskip 1em plus 0.5em
  minus 0.4em\relax USA: ACM Press/Addison-Wesley Publishing Co., 1999, p.
  335–342. [Online]. Available: \url{https://doi.org/10.1145/311535.311580}
\BIBentrySTDinterwordspacing

\bibitem{breen2001level}
\BIBentryALTinterwordspacing
D.~E. Breen and R.~T. Whitaker, ``{A Level-Set Approach for the Metamorphosis
  of Solid Models},'' in \emph{ACM SIGGRAPH 99 Conference Abstracts and
  Applications}, ser. SIGGRAPH '99.\hskip 1em plus 0.5em minus 0.4em\relax New
  York, NY, USA: Association for Computing Machinery, 1999, p. 228. [Online].
  Available: \url{https://doi.org/10.1145/311625.312113}
\BIBentrySTDinterwordspacing

\bibitem{bansal2018lie}
\BIBentryALTinterwordspacing
S.~Bansal and A.~Tatu, ``{Lie Bodies Based 3D Shape Morphing and
  Interpolation},'' in \emph{Proceedings of the 15th ACM SIGGRAPH European
  Conference on Visual Media Production}, ser. CVMP '18.\hskip 1em plus 0.5em
  minus 0.4em\relax New York, NY, USA: Association for Computing Machinery,
  2018. [Online]. Available: \url{https://doi.org/10.1145/3278471.3278477}
\BIBentrySTDinterwordspacing

\bibitem{kilian2007geometric}
\BIBentryALTinterwordspacing
M.~Kilian, N.~J. Mitra, and H.~Pottmann, ``{Geometric Modeling in Shape
  Space},'' \emph{ACM Trans. Graph.}, vol.~26, no.~3, p. 64–es, jul 2007.
  [Online]. Available: \url{https://doi.org/10.1145/1276377.1276457}
\BIBentrySTDinterwordspacing

\bibitem{bao2005physically}
\BIBentryALTinterwordspacing
Y.~Bao, X.~Guo, and H.~Qin, ``{Physically Based Morphing of Point-Sampled
  Surfaces},'' \emph{Computer Animation and Virtual Worlds}, vol.~16, no. 3-4,
  pp. 509--518, 2005. [Online]. Available:
  \url{https://doi.org/10.1002/cav.100}
\BIBentrySTDinterwordspacing

\bibitem{muller2004point}
\BIBentryALTinterwordspacing
M.~M\"{u}ller, R.~Keiser, A.~Nealen, M.~Pauly, M.~Gross, and M.~Alexa, ``{Point
  Based Animation of Elastic, Plastic and Melting Objects},'' in
  \emph{Proceedings of the 2004 ACM SIGGRAPH/Eurographics Symposium on Computer
  Animation}, ser. SCA '04.\hskip 1em plus 0.5em minus 0.4em\relax Goslar, DEU:
  Eurographics Association, 2004, p. 141–151. [Online]. Available:
  \url{https://doi.org/10.1145/1028523.1028542}
\BIBentrySTDinterwordspacing

\bibitem{dobashi2015simple}
\BIBentryALTinterwordspacing
Y.~Dobashi, T.~Tani, S.~Sato, and T.~Yamamoto, ``{A Simple Method for Morphing
  Smoke},'' in \emph{MATHEMATICAL PROGRESS IN EXPRESSIVE IMAGE SYNTHESIS},
  2015. [Online]. Available: \url{https://hdl.handle.net/2324/1546881}
\BIBentrySTDinterwordspacing

\bibitem{ludwig20153d}
\BIBentryALTinterwordspacing
M.~Ludwig, S.~Berrier, M.~Tetzlaff, and G.~Meyer, ``{3D Shape and Texture
  Morphing Using 2D Projection and Reconstruction},'' \emph{Comput. Graph.},
  vol.~51, no.~C, p. 146–156, oct 2015. [Online]. Available:
  \url{https://doi.org/10.1016/j.cag.2015.05.005}
\BIBentrySTDinterwordspacing

\bibitem{2000-ARAPShapeInterpolation}
\BIBentryALTinterwordspacing
M.~Alexa, D.~Cohen-Or, and D.~Levin, ``{A}s-{R}igid-{A}s-{P}ossible {S}hape
  {I}nterpolation,'' in \emph{Proceedings of the 27th Annual Conference on
  Computer Graphics and Interactive Techniques}, ser. SIGGRAPH '00.\hskip 1em
  plus 0.5em minus 0.4em\relax USA: ACM Press/Addison-Wesley Publishing Co.,
  2000, p. 157–164. [Online]. Available:
  \url{https://doi.org/10.1145/344779.344859}
\BIBentrySTDinterwordspacing

\bibitem{2004-ActualMorphing}
\BIBentryALTinterwordspacing
S.-M. Hu, C.-F. Li, and H.~Zhang, ``{A}ctual {M}orphing: {A} {P}hysics-{B}ased
  {A}pproach to {B}lending,'' in \emph{Proceedings of the Ninth ACM Symposium
  on Solid Modeling and Applications}, ser. SM '04.\hskip 1em plus 0.5em minus
  0.4em\relax Goslar, DEU: Eurographics Association, 2004, p. 309–314.
  [Online]. Available: \url{https://dl.acm.org/doi/abs/10.5555/1217875.1217924}
\BIBentrySTDinterwordspacing

\bibitem{2020-HamiltonianShapeInterpolation}
\BIBentryALTinterwordspacing
M.~Eisenberger and D.~Cremers, ``{H}amiltonian {D}ynamics for {R}eal-{W}orld
  {S}hape {I}nterpolation,'' \emph{Co{RR}}, vol. abs/2004.05199, 2020.
  [Online]. Available: \url{https://arxiv.org/abs/2004.05199}
\BIBentrySTDinterwordspacing

\bibitem{2005-PoissonShapeInterpolation}
\BIBentryALTinterwordspacing
D.~Xu, H.~Zhang, Q.~Wang, and H.~Bao, ``{P}oisson {S}hape {I}nterpolation,'' in
  \emph{Proceedings of the 2005 ACM Symposium on Solid and Physical Modeling},
  ser. SPM '05.\hskip 1em plus 0.5em minus 0.4em\relax New York, NY, USA:
  {A}ssociation for {C}omputing {M}achinery, 2005, p. 267–274. [Online].
  Available: \url{https://doi.org/10.1145/1060244.1060274}
\BIBentrySTDinterwordspacing

\bibitem{2011-ARAPSurfaceMorphing}
\BIBentryALTinterwordspacing
Y.-S. Liu, H.-B. Yan, and R.~R. Martin, ``{A}s-{R}igid-{A}s-{P}ossible
  {S}urface {M}orphing.'' [Online]. Available:
  \url{https://doi.org/10.1007/s11390-011-1154-3}
\BIBentrySTDinterwordspacing

\bibitem{2013-DataDrivenShapeMorphing}
\BIBentryALTinterwordspacing
L.~Gao, Y.~Lai, Q.~Huang, and S.~Hu, ``{A} {D}ata-{D}riven {A}pproach to
  {R}ealistic {S}hape {M}orphing,'' \emph{Comput. Graph. Forum}, vol.~32,
  no.~2, pp. 449--457, 2013. [Online]. Available:
  \url{https://doi.org/10.1111/cgf.12065}
\BIBentrySTDinterwordspacing

\bibitem{yang2021geometry}
\BIBentryALTinterwordspacing
G.~Yang, S.~Belongie, B.~Hariharan, and V.~Koltun, ``{Geometry Processing with
  Neural Fields},'' in \emph{Advances in Neural Information Processing
  Systems}, M.~Ranzato, A.~Beygelzimer, Y.~Dauphin, P.~Liang, and J.~W.
  Vaughan, Eds., vol.~34.\hskip 1em plus 0.5em minus 0.4em\relax Curran
  Associates, Inc., 2021, pp. 22\,483--22\,497. [Online]. Available:
  \url{https://proceedings.neurips.cc/paper_files/paper/2021/file/bd686fd640be98efaae0091fa301e613-Paper.pdf}
\BIBentrySTDinterwordspacing

\bibitem{sharpdeep}
\BIBentryALTinterwordspacing
F.~Gibou, D.~Hyde, and R.~Fedkiw, ``{Sharp Interface Approaches and Deep
  Learning Techniques for Multiphase Flows},'' \emph{Journal of Computational
  Physics}, vol. 380, pp. 442--463, 2019. [Online]. Available:
  \url{https://www.sciencedirect.com/science/article/pii/S0021999118303371}
\BIBentrySTDinterwordspacing

\bibitem{1988-SpacetimeConstraints}
\BIBentryALTinterwordspacing
A.~Witkin and M.~Kass, ``{Spacetime Constraints},'' \emph{SIGGRAPH Comput.
  Graph.}, vol.~22, no.~4, p. 159–168, 6 1988. [Online]. Available:
  \url{https://doi.org/10.1145/378456.378507}
\BIBentrySTDinterwordspacing

\bibitem{2020-ManipulateAmorphousMaterialsRL}
\BIBentryALTinterwordspacing
Y.~Zhang, W.~Yu, C.~K. Liu, C.~Kemp, and G.~Turk, ``{L}earning to {M}anipulate
  {A}morphous {M}aterials,'' \emph{ACM Trans. Graph.}, vol.~39, no.~6, 11 2020.
  [Online]. Available: \url{https://doi.org/10.1145/3414685.3417868}
\BIBentrySTDinterwordspacing

\bibitem{coros2021differentiable}
\BIBentryALTinterwordspacing
S.~Coros, M.~Macklin, B.~Thomaszewski, and N.~Th\"{u}rey, ``{Differentiable
  Simulation},'' in \emph{SIGGRAPH Asia 2021 Courses}, ser. SA '21.\hskip 1em
  plus 0.5em minus 0.4em\relax New York, NY, USA: Association for Computing
  Machinery, 2021. [Online]. Available:
  \url{https://doi.org/10.1145/3476117.3483433}
\BIBentrySTDinterwordspacing

\bibitem{2020-DiffPhysicsSim}
\BIBentryALTinterwordspacing
J.~Liang and M.~C. Lin, ``{D}ifferentiable {P}hysics {S}imulation,'' in
  \emph{ICLR 2020 Workshop on Integration of Deep Neural Models and
  Differential Equations}, 2019. [Online]. Available:
  \url{https://openreview.net/forum?id=p-SG2KFY2}
\BIBentrySTDinterwordspacing

\bibitem{2018-ChainQueen}
\BIBentryALTinterwordspacing
Y.~Hu, J.~Liu, A.~Spielberg, J.~B. Tenenbaum, W.~T. Freeman, J.~Wu, D.~Rus, and
  W.~Matusik, ``{C}hainqueen: {A} {R}eal-{T}ime {D}ifferentiable {P}hysical
  {S}imulator for {S}oft {R}obotics,'' in \emph{2019 International Conference
  on Robotics and Automation (ICRA)}, 2019, pp. 6265--6271. [Online].
  Available: \url{https://doi.org/10.1109/ICRA.2019.8794333}
\BIBentrySTDinterwordspacing

\bibitem{2019-DiffTaichi}
\BIBentryALTinterwordspacing
Y.~Hu, L.~Anderson, T.~Li, Q.~Sun, N.~Carr, J.~Ragan{-}Kelley, and F.~Durand,
  ``{DiffTaichi: Differentiable Programming for Physical Simulation},''
  \emph{CoRR}, vol. abs/1910.00935, 2019. [Online]. Available:
  \url{http://arxiv.org/abs/1910.00935}
\BIBentrySTDinterwordspacing

\bibitem{2019-DiffCloth}
\BIBentryALTinterwordspacing
J.~Liang, M.~Lin, and V.~Koltun, ``{Differentiable Cloth Simulation for Inverse
  Problems},'' in \emph{Advances in Neural Information Processing Systems},
  H.~Wallach, H.~Larochelle, A.~Beygelzimer, F.~d\textquotesingle
  Alch\'{e}-Buc, E.~Fox, and R.~Garnett, Eds., vol.~32.\hskip 1em plus 0.5em
  minus 0.4em\relax Curran Associates, Inc., 2019. [Online]. Available:
  \url{https://proceedings.neurips.cc/paper_files/paper/2019/file/28f0b864598a1291557bed248a998d4e-Paper.pdf}
\BIBentrySTDinterwordspacing

\bibitem{2022-DiffSoftMultiBody}
\BIBentryALTinterwordspacing
Y.-L. Qiao, J.~Liang, V.~Koltun, and M.~C. Lin, ``{Differentiable Simulation of
  Soft Multi-body Systems},'' 2022. [Online]. Available:
  \url{https://doi.org/10.48550/arXiv.2205.01758}
\BIBentrySTDinterwordspacing

\bibitem{2021-PODS}
\BIBentryALTinterwordspacing
M.~A.~Z. Mora, M.~Peychev, S.~Ha, M.~Vechev, and S.~Coros, ``{PODS: Policy
  Optimization via Differentiable Simulation},'' in \emph{Proceedings of the
  38th International Conference on Machine Learning}, ser. Proceedings of
  Machine Learning Research, M.~Meila and T.~Zhang, Eds., vol. 139.\hskip 1em
  plus 0.5em minus 0.4em\relax PMLR, 7 2021, pp. 7805--7817. [Online].
  Available: \url{https://proceedings.mlr.press/v139/mora21a.html}
\BIBentrySTDinterwordspacing

\bibitem{2022-DiffSimRL}
\BIBentryALTinterwordspacing
J.~Xu, V.~Makoviychuk, Y.~Narang, F.~Ramos, W.~Matusik, A.~Garg, and
  M.~Macklin, ``{Accelerated Policy Learning with Parallel Differentiable
  Simulation},'' 2022. [Online]. Available:
  \url{https://doi.org/10.48550/arXiv.2204.07137}
\BIBentrySTDinterwordspacing

\bibitem{sulsky1994particle}
\BIBentryALTinterwordspacing
D.~Sulsky, Z.~Chen, and H.~Schreyer, ``{A Particle Method for History-Dependent
  Materials},'' \emph{{Computer Methods in Applied Mechanics and Engineering}},
  vol. 118, no.~1, pp. 179--196, 1994. [Online]. Available:
  \url{https://www.sciencedirect.com/science/article/pii/0045782594901120}
\BIBentrySTDinterwordspacing

\bibitem{jiang2016material}
\BIBentryALTinterwordspacing
C.~Jiang, C.~Schroeder, J.~Teran, A.~Stomakhin, and A.~Selle, ``{The Material
  Point Method for Simulating Continuum Materials},'' in \emph{ACM SIGGRAPH
  2016 Courses}, ser. SIGGRAPH '16.\hskip 1em plus 0.5em minus 0.4em\relax New
  York, NY, USA: Association for Computing Machinery, 2016. [Online].
  Available: \url{https://doi.org/10.1145/2897826.2927348}
\BIBentrySTDinterwordspacing

\bibitem{2018-MLSMPM}
\BIBentryALTinterwordspacing
Y.~Hu, Y.~Fang, Z.~Ge, Z.~Qu, Y.~Zhu, A.~Pradhana, and C.~Jiang, ``{A} {M}oving
  {L}east {S}quares {M}aterial {P}oint {M}ethod with {D}isplacement
  {D}iscontinuity and {T}wo-way {R}igid {B}ody {C}oupling,'' \emph{ACM Trans.
  Graph.}, vol.~37, no.~4, jul 2018. [Online]. Available:
  \url{https://doi.org/10.1145/3197517.3201293}
\BIBentrySTDinterwordspacing

\bibitem{10.1145/3130800.3130820}
\BIBentryALTinterwordspacing
Y.~Zhu, J.~Popovi\'{c}, R.~Bridson, and D.~M. Kaufman, ``Planar {I}nterpolation
  with {E}xtreme {D}eformation, {T}opology {C}hange and {D}ynamics,'' \emph{ACM
  Trans. Graph.}, vol.~36, no.~6, 11 2017. [Online]. Available:
  \url{https://doi.org/10.1145/3130800.3130820}
\BIBentrySTDinterwordspacing

\bibitem{2019-cdmpm}
\BIBentryALTinterwordspacing
J.~Wolper, Y.~Fang, M.~Li, J.~Lu, M.~Gao, and C.~Jiang, ``{CD-MPM}: {C}ontinuum
  {D}amage {M}aterial {P}oint {M}ethods for {D}ynamic {F}racture {A}nimation,''
  \emph{ACM Trans. Graph.}, vol.~38, no.~4, jul 2019. [Online]. Available:
  \url{https://doi.org/10.1145/3306346.3322949}
\BIBentrySTDinterwordspacing

\bibitem{fang2020iq}
\BIBentryALTinterwordspacing
Y.~Fang, Z.~Qu, M.~Li, X.~Zhang, Y.~Zhu, M.~Aanjaneya, and C.~Jiang, ``{IQ-MPM:
  An Interface Quadrature Material Point Method for Non-Sticky Strongly Two-Way
  Coupled Nonlinear Solids and Fluids},'' \emph{ACM Trans. Graph.}, vol.~39,
  no.~4, aug 2020. [Online]. Available:
  \url{https://doi.org/10.1145/3386569.3392438}
\BIBentrySTDinterwordspacing

\bibitem{bardenhagen2004generalized}
\BIBentryALTinterwordspacing
E.~M.~K. S.~G.~Bardenhagen, ``{The Generalized Interpolation Material Point
  Method},'' \emph{Computer Modeling in Engineering \& Sciences}, vol.~5,
  no.~6, pp. 477--496, 2004. [Online]. Available:
  \url{http://www.techscience.com/CMES/v5n6/33378}
\BIBentrySTDinterwordspacing

\bibitem{solowski2021material}
\BIBentryALTinterwordspacing
W.~T. Sołowski, M.~Berzins, W.~M. Coombs, J.~E. Guilkey, M.~Möller, Q.~A.
  Tran, T.~Adibaskoro, S.~Seyedan, R.~Tielen, and K.~Soga, ``Chapter two -
  material point method: Overview and challenges ahead,'' ser. Advances in
  Applied Mechanics, S.~P. Bordas and D.~S. Balint, Eds.\hskip 1em plus 0.5em
  minus 0.4em\relax Elsevier, 2021, vol.~54, pp. 113--204. [Online]. Available:
  \url{https://www.sciencedirect.com/science/article/pii/S0065215620300120}
\BIBentrySTDinterwordspacing

\bibitem{de2020material}
\BIBentryALTinterwordspacing
A.~{de Vaucorbeil}, V.~P. Nguyen, S.~Sinaie, and J.~Y. Wu, ``{Chapter Two -
  Material Point Method After 25 Years: Theory, Implementation, and
  Applications},'' ser. Advances in Applied Mechanics, S.~P. Bordas and D.~S.
  Balint, Eds.\hskip 1em plus 0.5em minus 0.4em\relax Elsevier, 2020, vol.~53,
  pp. 185--398. [Online]. Available:
  \url{https://www.sciencedirect.com/science/article/pii/S0065215619300146}
\BIBentrySTDinterwordspacing

\bibitem{jiang2017angular}
\BIBentryALTinterwordspacing
C.~Jiang, C.~Schroeder, and J.~Teran, ``{A}n {A}ngular {M}omentum {C}onserving
  {A}ffine-{P}article-in-{C}ell {M}ethod,'' \emph{Journal of {C}omputational
  {P}hysics}, vol. 338, pp. 137--164, 2017. [Online]. Available:
  \url{https://www.sciencedirect.com/science/article/pii/S0021999117301535}
\BIBentrySTDinterwordspacing

\bibitem{2015-APIC}
\BIBentryALTinterwordspacing
C.~Jiang, C.~Schroeder, A.~Selle, J.~Teran, and A.~Stomakhin, ``{The Affine
  Particle-in-Cell Method},'' \emph{ACM Trans. Graph.}, vol.~34, no.~4, 7 2015.
  [Online]. Available: \url{https://doi.org/10.1145/2766996}
\BIBentrySTDinterwordspacing

\bibitem{johnson2023software}
\BIBentryALTinterwordspacing
D.~Johnson, T.~Maxfield, Y.~Jin, and R.~Fedkiw, ``{Software-Based Automatic
  Differentiation is Flawed},'' \emph{arXiv preprint arXiv:2305.03863}, 2023.
  [Online]. Available: \url{https://doi.org/10.48550/arXiv.2305.03863}
\BIBentrySTDinterwordspacing

\bibitem{kingma2014adam}
D.~P. Kingma, ``Adam: A method for stochastic optimization,'' \emph{arXiv
  preprint arXiv:1412.6980}, 2014.

\bibitem{2012-FixedCoratedElasticty}
\BIBentryALTinterwordspacing
A.~Stomakhin, R.~Howes, C.~Schroeder, and J.~M. Teran, ``Energetically
  {C}onsistent {I}nvertible {E}lasticity,'' in \emph{Proceedings of the ACM
  SIGGRAPH/Eurographics Symposium on Computer Animation}, ser. SCA '12.\hskip
  1em plus 0.5em minus 0.4em\relax Goslar, DEU: {Eurographics Association},
  2012, p. 25–32. [Online]. Available:
  \url{https://dl.acm.org/doi/10.5555/2422356.2422361}
\BIBentrySTDinterwordspacing

\bibitem{gast2016implicit}
\BIBentryALTinterwordspacing
T.~F. Gast, C.~Fu, C.~Jiang, and J.~Teran, ``{Implicit-shifted Symmetric QR
  Singular Value Decomposition of 3x3 Matrices},'' 2016. [Online]. Available:
  \url{https://api.semanticscholar.org/CorpusID:11952786}
\BIBentrySTDinterwordspacing

\bibitem{10.1145/3606037.3606840}
\BIBentryALTinterwordspacing
M.~Xu and D.~I. Levin, ``{Deformation Gradient Control of Physically Simulated
  Amorphous Solids},'' in \emph{Proceedings of the ACM SIGGRAPH/Eurographics
  Symposium on Computer Animation}, ser. SCA '23.\hskip 1em plus 0.5em minus
  0.4em\relax New York, NY, USA: Association for Computing Machinery, 2023.
  [Online]. Available: \url{https://doi.org/10.1145/3606037.3606840}
\BIBentrySTDinterwordspacing

\bibitem{2011-ShapeCorrespondenceSurvey}
\BIBentryALTinterwordspacing
O.~v. Kaick, H.~Zhang, G.~Hamarneh, and D.~Cohen‐Or, ``{A Survey on Shape
  Correspondence},'' \emph{Computer Graphics Forum}, 2011. [Online]. Available:
  \url{http://dx.doi.org/10.1111/j.1467-8659.2011.01884.x}
\BIBentrySTDinterwordspacing

\bibitem{hochreiter1998vanishing}
S.~Hochreiter, ``The vanishing gradient problem during learning recurrent
  neural nets and problem solutions,'' \emph{International Journal of
  Uncertainty, Fuzziness and Knowledge-Based Systems}, vol.~6, no.~02, pp.
  107--116, 1998.

\bibitem{polyscope}
\BIBentryALTinterwordspacing
N.~Sharp \emph{et~al.}, ``{Polyscope},'' 2019. [Online]. Available:
  \url{www.polyscope.run}
\BIBentrySTDinterwordspacing

\bibitem{wang2020massively}
\BIBentryALTinterwordspacing
X.~Wang, Y.~Qiu, S.~R. Slattery, Y.~Fang, M.~Li, S.-C. Zhu, Y.~Zhu, M.~Tang,
  D.~Manocha, and C.~Jiang, ``{A Massively Parallel and Scalable Multi-GPU
  Material Point Method},'' \emph{ACM Trans. Graph.}, vol.~39, no.~4, aug 2020.
  [Online]. Available: \url{https://doi.org/10.1145/3386569.3392442}
\BIBentrySTDinterwordspacing

\end{thebibliography}

\begin{IEEEbiography}[{\includegraphics[width=1in,height=1.25in,clip,keepaspectratio]{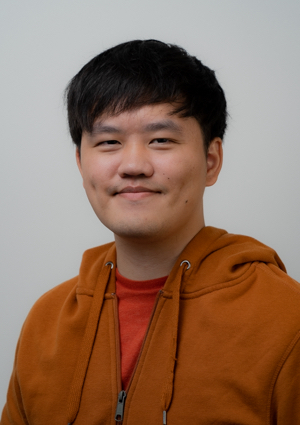}}]{Michael Xu}
received the B.S.\ in in Engineering Science at the University of Toronto. He is currently pursuing a Ph.D.\ in Computing Science at Simon Fraser University. He is interested in applying machine learning and optimization techniques to physics-based character control.
\end{IEEEbiography}

\begin{IEEEbiography}[{\includegraphics[width=1in,height=1.25in,clip,keepaspectratio]{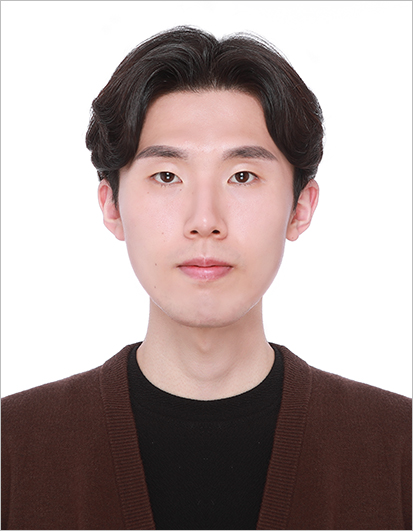}}]{Chang-Yong Song}
received the M.S.\ degree from Korea University in 2023. He is currently pursuing a Ph.D.\ at Vanderbilt University under the supervision of Dr.\ David Hyde. His research interests include physically-based animation for deformable bodies, with an additional focus on machine learning. His goal is to create compelling animations that leave a lasting impression on people.
\end{IEEEbiography}

\begin{IEEEbiographynophoto}{David Levin}
Photograph and biography not available at the time of publication.
\end{IEEEbiographynophoto}

\begin{IEEEbiographynophoto}{David Hyde}
(Member, IEEE) Photograph and biography not available at the time of publication.
\end{IEEEbiographynophoto}

\newpage
\clearpage

 \appendix [MLS-MPM]
\section{*MLS-MPM}
\begin{table}[ht!]
    \caption{\textmd{A table of variables used in our {MPM} simulation formulation.}}
    \centering
    \begin{tabular}{ccc}
    \toprule
        Symbol & Type & Meaning \\ 
    \midrule
    $\rho$ & scalar & density (kg/$m^3$)  \\ 
    $\lambda$ & scalar & material parameter \\
    $\mu$ & scalar & material parameter \\
    $\zeta$ & scalar & damping coefficient\\ 
    $\gamma$ & scalar & smoothing factor\\
    $\Delta t$ & scalar & time step size  \\
    $\Delta x$ & scalar & grid node spacing \\
    \midrule	
    $\mathbf{x}_{p}$ & vector & position \\
    $m_{p}$ & scalar & mass \\
    $V_{0}$ & scalar & initial volume \\
    $\mathbf{v}_{p}$ & vector & velocity \\
    \midrule	
    $\mathbf{C}_{p}$ & matrix & affine velocity field \\
    $\mathbf{P}_{p}$ & matrix & PK1 stress tensor \\
    $\mathbf{R}$ & matrix & rotation matrix  \\
    $\mathbf{F}_{p}$ & matrix & deformation gradient  \\
    $\mathbf{\tilde{F}}_{p}$ & matrix & control deformation gradient \\
    \midrule	
    $\mathbf{x}_{i}$ & vector & position \\
    $m_{i}$ & scalar & mass  \\
    $\mathbf{v}_{i}$ & vector & velocity \\
    $\mathbf{p}_{i}$ & vector & momentum \\
    $\mathbf{f}_{\text{ext}}$ & vector & external force  \\
    $N$ & scalar & cubic B-spline function \\
    \bottomrule
    \end{tabular}
    \vspace{1mm}
    \label{tab:mpm-syms}
    \\
    \footnotemark{The subscripts index into the domain of the symbol, where $p$ indexes into material points and $i$ indexes into background grid nodes.}
\end{table}

\renewcommand{\theequation}{A\arabic{equation}} 
\setcounter{equation}{0}  

Here we describe our implementation of the {MLS-MPM} algorithm, as well as the minor modifications we made to the algorithm and the ensuing gradients. 
For an in-depth introduction to {MPM}, we refer the reader to the {MPM} course notes by Jiang et al. \cite{jiang2016material}.
We also refer to Hu et al. \cite{2018-ChainQueen}, which derived the analytical gradients for the unmodified {MLS-MPM} algorithm.
Upon publication, we will upload our code on GitHub under a permissive license, so that the modified algorithm and gradients we used will be readily available.

The first modification we make to the standard algorithm is to introduce a damping coefficient $\zeta$, which is helpful for running explicit {MPM} simulations with large timesteps.
We used $\zeta = 0.5$ for all our examples.
We incorporate this term when interpolating particle momenta to momentum on grid nodes, see Equation \ref{eqn:gridp}.

The second change we make is the introduction of a control deformation gradient variable $\mathbf{\tilde{F}}_{p}$.
Section \ref{sec:control} discusses the details of using this term in the morphing algorithm.

\newcommand{\PK}[2]{\mathbf{P}_{#1} ^{#2} }

\newcommand{\F}[2]{\mathbf{F}_{#1} ^{#2} }

The third change was implemented to create a smoother transition in the deformation gradient changes. This was achieved by using a smoothing factor $\gamma$ to apply a linear combination of the previous and next gradients, i.e. $\textbf{F}^{n+1} = (1 - \gamma)\textbf{F}^{n+1} + \gamma\textbf{F}^{n}$. This approach helped to eliminate oscillations in the simulation (see Section \ref{sec:examples}).

With these three changes, the modified {MLS-MPM} algorithm we use proceeds as follows:

\newcommand{\Fc}[2]{\tilde{\mathbf{F}}_{#1} ^{#2} }

\newcommand{\dip}[1]{\mathbf{d}_{i p} ^{#1} }
\newcommand{\x}[2]{\mathbf{x}_{#1} ^{#2} }
\newcommand{\wip}[1]{w_{i p} ^{#1} }
\newcommand{\gridp}[1]{\mathbf{p}_i ^{#1} }
\newcommand{\dt}{\Delta t}
\newcommand{\damp}{\zeta}
\newcommand{\dx}{\Delta x}
\newcommand{\C}[2]{\mathbf{C}_{#1} ^{#2} }
\newcommand{\G}[2]{\mathbf{G}_{#1} ^{#2} }
\newcommand{\initialvol}[1]{V_{#1} ^0 }
\newcommand{\vel}[2]{\mathbf{v}_{#1} ^{#2} }
\newcommand{\fext}{\mathbf{f}_{\text{ext}} }

\emph{P1.} The first step of {MPM} is to compute each particle's Piola-Kirchoff stress tensor, which is a function of the deformation gradient $\mathbf{F}$ and depends on the constitutive model.
We use the fixed-corotated linear elastic model \cite{2012-FixedCoratedElasticty, jiang2016material}, which makes use of the determinant $J$ of $\mathbf{F}$ and the polar decomposition \cite{gast2016implicit} of $\mathbf{F}$ into its rotation matrix $\mathbf{R}$ and its symmetric scaling matrix $\mathbf{S}$:

    \begin{equation}
    \PK{ }{ }(\mathbf{F}) = 2 \mu (\mathbf{F} - \mathbf{R}) + \lambda  (J - 1)  J  \mathbf{F}^{-T} .
    \end{equation}
    
The total deformation gradient on a material point at a given timestep is the sum of the time-evolved deformation gradient $\F{p}{n}$ and the control parameter $\Fc{p}{n}$.
In all instances where the time-evolved deformation gradient is used in the normal MPM algorithm, we instead use the sum of it and the control parameter $\Fc{p}{n}$.
Using this sum allows the loss function to have identical gradients with respect to $\F{p}{n}$ and $\Fc{p}{n}$.
It also continues to allow the deformation gradient $\F{p}{n}$ to time evolve rather than be overwritten by a control deformation gradient, which can be seen in the P2 step.
Accordingly, we compute the per-particle stress by evaluating the prior equation at $\F{p}{n} + \Fc{p}{n}$, i.e., $\PK{p}{n} = \PK{ }{ } (\F{p}{n} + \Fc{p}{n})$.

\emph{P2G.} The second step of MPM is to interpolate particle mass and momentum onto the grid nodes, which is the precursor step to calculating updated grid node velocities.
We use APIC \cite{2015-APIC, jiang2017angular} for this transfer, which requires an affine velocity field $\C{p}{ }$ stored as a matrix on each of our material points.
For ease of exposition, we define helper variables $\dip{n}$, $\wip{n}$, and $\G{p}{n}$ used in the transfers:
\begin{align}
    \dip{n} & = \x{i}{ } - \x{p}{n} , \\
    \wip{n} & = N(\dip{n}) , \\
    \G{p}{n} & = \frac{-3}{\dx^2} \dt \initialvol{p} \PK{ }{ }\left( \F{p}{n}+\Fc{p}{n} \right) ^{T} + m_p \C{p}{n} , \label{eqn:gpn}
\end{align}
where Equation \ref{eqn:gpn} uses the moving least squares (MLS) force discretization \cite{2018-ChainQueen}. 
With these variables, we may express the interpolated grid node masses as
\begin{equation}
    m_i ^n =  \sum_p w_{i p} m_p,
\end{equation}
and the interpolated grid node momenta as
\begin{equation}
    \gridp{n} = \sum_p \wip{n}(m_p \vel{p}{n} \cdot(1-\damp \dt) + \G{p}{n} \dip{n}) .
    \label{eqn:gridp}
\end{equation}
\emph{G.} The third step of {MPM} is to calculate nodal velocities by dividing the nodal momentum by the nodal mass. Accelerations from external forces such as gravity are also applied here:
\begin{equation}
    \vel{i}{n} = \frac{1}{m_i ^n}(\gridp{n} + \fext \dt) .
\end{equation}

\emph{G2P.} In the previous step, the velocity of the grid nodes were calculated by interpolating nearby material point properties and then updated by external forces such as gravity. 
Updated velocities are then interpolated back to nearby material points:
\begin{equation}
    \vel{p}{n+1} = \sum_i \wip{n} \vel{i}{n} .
\end{equation}

Updated material point affine velocity matrices \cite{2015-APIC, jiang2017angular} must also be computed from the grid nodes. 
The equation we provide is specific to using cubic B-spline interpolation functions:
\begin{equation}
    \C{p}{n+1} = \frac{3}{\dx^2} \sum_i \wip{n} \vel{i}{n} (\dip{n}) ^{T} .
\end{equation}

\textit{P2.} The final step of an {MPM} timestep is to update the material point positions and deformation gradients, where we use a specified tolerance $\gamma$ to control the deformation update:
\begin{align}
    \x{p}{n+1} & = \x{p}{n} + \dt \vel{p}{n+1} \\
    \F{p}{n+\frac{1}{2}} & = (\mathbf{I} + \dt \C{p}{n+1}) (\F{p}{n} + \Fc{p}{n+1})\\
    \beta &= u(||\F{p}{n+\frac{1}{2}} - \F{p}{n}|| - \gamma)\\
    \F{p}{n+1} &= (1 - \beta) \F{p}{n+\frac{1}{2}} + \beta \F{p}{n}
\end{align}
where $u$ is the unit step function. When $\beta = 1$, i.e. when the deformation change exceeds our allowed tolerance $\gamma$, the deformation update is equivalent to overwriting the control deformation gradient $\Fc{p}{n+1}$ such that $\Fc{p}{n+1} = (\mathbf{I} + \dt \C{p}{n+1})^{-1} (\F{p}{n} - (\mathbf{I} + \dt \C{p}{n+1}) \F{p}{n})$.

\end{document}